\documentclass[prx, letterpaper, superscriptaddress, amsmath, amssymb, amssymb, reprint, floatfix]{revtex4-2}

\usepackage{units}
\usepackage{xfrac}
\usepackage{xcolor}
\usepackage{graphicx}
\usepackage{dcolumn}
\usepackage{bm}
\usepackage[T1]{fontenc}

\begin{document}

\preprint{APS/123-QED}

\title{Measurement of cryoelectronics heating \\ using a local quantum dot thermometer in silicon}

\author{Mathieu de Kruijf}
 \email{mathieu@quantummotion.tech}
\affiliation{
 Quantum Motion, 9 Sterling Way, London, 
 N7 9HJ, United Kingdom
}
\affiliation{%
 London Centre for Nanotechnology, University College London, 17-19 Gordon Street, London WC1H 0AH, United Kingdom
}%

\author{Grayson M. Noah}%
\affiliation{
 Quantum Motion, 9 Sterling Way, London, 
 N7 9HJ, United Kingdom
}%

\author{Alberto Gomez-Saiz}
\affiliation{
 Quantum Motion, 9 Sterling Way, London, 
 N7 9HJ, United Kingdom
}

\author{John J. L. Morton}
\affiliation{%
 London Centre for Nanotechnology, University College London,
 17-19 Gordon Street, London WC1H 0AH, United Kingdom
}%
\affiliation{
 Quantum Motion, 9 Sterling Way, London, N7 9HJ, United Kingdom
}%

\author{M. Fernando Gonzalez-Zalba}
\affiliation{
 Quantum Motion, 9 Sterling Way, London, N7 9HJ, United Kingdom
}%

\date{\today}

\begin{abstract}
Silicon technology offers the enticing opportunity for monolithic integration of quantum and classical electronic circuits. However, the power consumption levels of classical electronics may compromise the local chip temperature and hence the fidelity of qubit operations. Here, we utilize a quantum-dot-based thermometer embedded in an industry-standard silicon field-effect transistor (FET), to assess the local temperature increase produced by an active FET placed in close proximity. We study the impact of both static and dynamic operation regimes. When the FET is operated statically, we find a power budget of 45~nW at 100~nm separation whereas at 216~$\mu$m the power budget raises to 150~$\mu$W. When operated dynamically, we observe negligible temperature increase for the switch frequencies tested up to 10~MHz. Our work describes a method to accurately map out the available power budget at a distance from a solid-state quantum processor and indicate under which conditions cryoelectronics circuits may allow the operation of hybrid quantum-classical systems.

\end{abstract}

\maketitle

\section{\label{sec:Intro}Introduction}

Quantum computing based on spins in silicon quantum dots (QDs) is showing great progress towards a scalable quantum computing system~\cite{GonzalezZalba2021}. One- and two-qubit gates have been shown to perform above fault-tolerant thresholds for electron spins~\cite{Xue2022, Noiri2022, Mills2022}, simple instances of quantum error correction have been performed~\cite{Takeda2022} and a 6-qubit quantum processor has been demonstrated~\cite{Philips2022}. Besides, silicon-based quantum computing can benefit from industrial manufacturing techniques~\cite{Maurand2016, Zwerver2022, Camenzind2021} and monolithic integration with classical control and readout electronics~\cite{Schaal2019,Ruffino2022}. Monolithic integration compared to distributed quantum computing offers the benefits of ease of data flow, minimized latency, reduced system footprint and reduced cost~\cite{Reilly2019}. To this matter, understanding the impact of classical electronics located in close proximity to qubit devices is essential to learn under which conditions integration could be possible. 

Several studies have addressed the impact of classical control electronics on qubit fidelity from a theoretical~\cite{Dijk2019} and an experimental perspective~\cite{Xue2021,CBeltran2022,underwood2023using}. Furthermore, a number of 3~K cryogenic electronics peripherals have been demonstrated for the control~\cite{Bardin2019,Pauka2021,Park2021} and readout~\cite{Guevel2020, Prabowo2021,Ruffino2021} of solid-state qubits while some studies have attempted to integrate qubits and readout electronics on the same chip~\cite{Guevel2020_2, Ruffino2022}. The study of the impact of power dissipated by classical electronics has been primarily focused on investigations of self-heating~\cite{Triantopoulos2019, Hart2021} while a number of experiments have investigated the impact of operating qubits, and particularly spins in silicon, at elevated temperatures~\cite{Petit2020,Yang2020, undseth2023}. More recently, on-chip thermometry methods native to complementary metal-oxide-semiconductor (CMOS) industrial fabrication processes have been used to characterise the impact of static power dissipation~\cite{noah2023cmos}.

In this Article, we build on the latter work and focus on understanding the impact of active classical electronics, with their associated static and dynamic power dissipation, on the temperature of a monolithically integrated QD device, all fabricated using an industry-standard process. More particularly, we utilize electronic transport through the QD as a sensitive thermometer to detect the temperature increase associated with the different driving modes of a field-effect transistor (FET). We first present the thermometry principle in a QD device exhibiting a discrete density of states. We then explain the calibration procedure of the thermometer and discuss the mechanisms that limit the range of operation of the sensor. We then utilize the thermometer to quantify the power that substantially raises the local temperature under static bias conditions of the FET for two different QD-FET separations. Finally, we quantify the effect of dynamic power dissipation in the FET for switching rates of up to 10 MHz. Overall, we provide a methodology to map out the power budget at a distance from a target QD.

\section{QD THERMOMETRY AND HEATING ELEMENT}

\subsection{Device}

\begin{figure}
\includegraphics[width=0.5\textwidth]{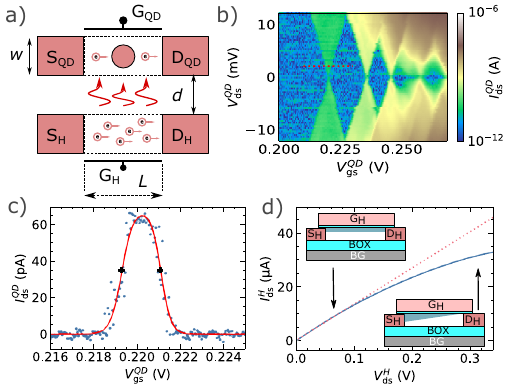}
\caption{\label{fig:1} Double FET. \textbf{(a)} A schematic of the device, where the top FET is used to form a QD and the bottom FET is used as a local heater element; channel width ($w$), gate lenght ($L$) and face-to-face distance ($d$) are indicated. \textbf{(b)} Coulomb blockade measurement for the QD in the top FET. The red dashed line indicates the  region where the IV curve  in (c) is taken. \textbf{(c)} IV curve at $V_\text{ds}=2$~mV including a fit to Eq.~\ref{eq:tophat} in red. The crosses indicate the points at which the electrochemical level of the QD aligns with the source and drain Fermi levels. \textbf{(d)} The drain-source current through the heater element as a function of drain-source voltage at $V_\text{gs}=0.8$~V. At high drain-source voltage, the FET goes into saturation and the current no longer follows a linear trend. The insets show schematically both operating regimes of the heater element.}
\end{figure}

For our experiments, we use two FETs manufactured using industry-standard fully-depleted silicon-on-insulator (FDSOI) technology and measure them in a dilution refrigerator with variable base temperature down to $T_\text{MXC}=20$~mK. We operate one of the FETs in the sub-threshold regime as a QD device and the other as a standard transistor in the inversion regime which we refer to as the heating element. In Fig.~\ref{fig:1}(a), we show a schematic of the system. The transistors have a physical gate length, $L=40$~nm, and a channel width, $w=80$~nm, and both share a common back gate with a ultra-thin buried oxide. Both transistors are separated by a SiO$_2$ shallow trench. We focus on two samples with channel edge-to-edge distance $d$, of 100~nm and 216~$\mu$m, respectively (see Appendix~\ref{app:methods}). 

First, we demonstrate the presence of a QD with discrete energy states in the channel of the top FET when operated in the sub-threshold region. We measure the drain-source current, $I_\text{ds}^\text{QD}$, as a function of the gate-source, $V_\text{gs}^\text{QD}$, and drain-source voltages, $V_\text{ds}^\text{QD}$, for a back gate voltage of $V_\text{bs}=2$~V. We find the characteristic features of Coulomb blockade: diamond-shaped regions where the current flow is substantially reduced, see Fig.~\ref{fig:1}(b)~\cite{Kouwenhoven2001}. We focus on the first Coulomb oscillation occurring at $V_\text{gs}^\text{QD}\approx 0.22$~V. A $V_\text{gs}^\text{QD}$ scan at constant $V_\text{ds}^\text{QD}=2$~mV (red line), reveals a top hat lineshape characteristic of a single discrete energy level (0D density of states) tunnel coupled to two reservoirs with a 3D density of states~\cite{Gonzalez-Zalba2015}, see Fig.~\ref{fig:1}(c). Here, we fit the data to a simple sequential single-electron tunneling expression,

\begin{equation}
    I_{\mathrm{ds}} = e\frac{\Gamma_{\mathrm{s}} \Gamma_{\mathrm{d}}}{\Gamma_{\mathrm{s}}+\Gamma_{\mathrm{d}}}, 
    \label{eq:tophat}
\end{equation}

\noindent in which the source(drain) reservoir to dot tunnel rates, $\Gamma_\text{s(d)}$, are defined as~\cite{Houten1992,Maradan2014},

\begin{equation}
    \Gamma_{\mathrm{s(d)}} = \frac{\Gamma_\text{0,s(d)}}{1+\exp{\left[ \mp e\alpha_\text{g} ( V_{\mathrm{gs}}^\text{QD}-V_{0,\text{s(d)}})/(k_\mathrm{B}T_\text{s(d)}) \right]} }.
    \label{eq:fermi}
\end{equation}

Here, $k_{\mathrm{B}}$ is the Boltzmann constant, $\Gamma_\text{0,s(d)}$ is the maximum source(drain) tunnel rate, and $T_\text{s(d)}$ the source(drain) reservoir temperatures. Further, the gate lever arm $\alpha_\text{g}$ --- the ratio between the gate and total capacitance of the QD --- is a calibration factor that translates $V_{\mathrm{gs}}^\text{QD}$ to an energy detuning between the QD electrochemical level and the Fermi level in the corresponding reservoir. The parameter $V_{\mathrm{0,s(d)}}$ is the gate-source voltage at which the QD and source(drain) Fermi level are aligned and can be used to extract the gate lever arm from a single IV trace as $\alpha_\text{g}=V_\text{ds}^\text{QD}/[V_{0,\text{d}}-V_{0,\text{s}}]$ and obtain $\alpha_\text{g}=0.91\pm0.10$; see Fig.~\ref{fig:1}(c). Equation~\ref{eq:fermi} is proportional to the Fermi-Dirac distribution and hence can be utilized to infer the electronic temperature in the corresponding reservoir. For $V_\text{ds}^\text{QD}>0$~V, we utilize the drain side (falling edge) of the lineshape for temperature extraction since it avoids the impact of excited states in the measurement bias window~\cite{Escott_2010} (see Appendix~\ref{app:Impact of excited states}).

Next, we characterize the heating FET which will be utilized to increase the chip temperature locally. In Fig.~\ref{fig:1}(d), we plot the drain-source current, $I_\text{ds}^\text{H}$, as a function of the drain-source voltage, $V_\text{ds}^\text{H}$, when the device gate-source bias is well above threshold, $V_\text{gs}^\text{H}=0.8$V. At low $V_\text{ds}^\text{H}$, we observe the linear regime of the transistor where the current increases proportionally to $V_\text{ds}^\text{H}$ and the inversion layer thickness is constant. However, at $V_\text{ds}^\text{H} > 0.1$~V, the IV curve deviates from the linear trend indicating that the FET is entering the saturation region. Here the inversion layer thickness decreases on the drain side as the system approaches pinch off. We utilize the heating FET to dissipate powers up to 2~$\mu$W.   

\subsection{Operation regime}

\begin{figure}
\includegraphics{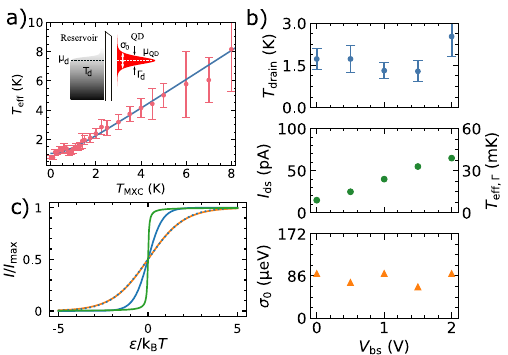}
\caption{\label{fig:2}Thermometry and transport regime. \textbf{(a)} Measured effective QD temperature vs MXC temperature after calibration through $\alpha_\text{g}$ (red dots) and fit to Eq.~\ref{eq:cal} (blue line).  \textbf{(inset)} Schematic of the three different physical mechanism that can affect the peak lineshape: Temperature of the reservoir ($T_\text{d}$) , lifetime broadening of the discrete state ($\Gamma_\text{d}$) and charge noise ($\sigma_0$). Here $\mu_\text{d}$ represents the Fermi level in the reservoir, and $\mu_\text{QD}$ represents the electrochemical level of the QD. \textbf{(b)} (Top) Back gate dependence of the QD effective temperature. (Middle) Maximum current through the top hat transition (left) and the associated equivalent temperature (right) as a function of the back gate potential considering equal source and drain tunnel rates, $I_\text{ds}=e\Gamma_\text{d}/2$.  (Bottom) Integrated charge noise over the bandwidth 0.005 to 0.1 Hz. \textbf{(c)} Calculated lineshapes for thermal broadening at $T=0.64$~K  (blue), lifetime broadening for 100~pA (green), charge noise broadening for $\sigma_0=100$~$\mu$V and $T=1.12$~K (orange) and overlapping thermal broadening at 1.3~K (dashed blue). The current is normalised to its maximum value $I_\text{max}$ and detuning to $T=1.3$~K. }
\end{figure}

To confirm the calibration of the QD temperature sensor, we record Coulomb blockade traces like in Fig.~\ref{fig:1}(c) for increasing mixing chamber temperature. We perform 20 independent measurements to which we fit Eq.~\ref{eq:tophat},  and plot the average and the standard deviation of the results for every step in MXC temperature in Fig.~\ref{fig:2}(a). We fit the data to the expression,

\begin{equation}
   T_\text{eff}= \sqrt{T_\text{MXC}^2+T_\text{0}^2},
   \label{eq:cal}
\end{equation}

that aims to capture the expected one-to-one dependence between the effective sensor temperature $T_\text{eff}$ and the MXC temperature as well as the saturation temperature $T_\text{0}$ at low $T_\text{MXC}$. We observe three regions in the data: (i) The region  between 1.5~K~and 6~K shows a good one-to-one match between the sensor's effective temperature $T_\text{eff}$ and $T_\text{MXC}$ (this region can be utilized for local thermometry); (ii) the region $T_\text{MXC}\gtrsim 6$~K, where the one-to-one relation becomes inaccurate as the the Fermi level broadening approaches the drain-source excitation ($3.5k_\text{B}T_\text{MXC}\approx eV_\text{ds}^\text{QD})$; and (iii) the region  $T_\text{MXC}\lesssim 1.5$~K when the temperature reading start to saturate as it approaches $T_0=1.03\pm0.05$~K (extracted from the fit to Eq.~\ref{eq:cal}). Below this temperature, the QD cannot be used as a sensitive thermometer. 

We envision three mechanisms leading to the saturation of the effective temperature well above the base temperature of the dilution refrigerator: (i) electron-phonon decoupling characterised by an electron temperature $T_\text{elect}$~\cite{Iftikhar2016}, (ii) lifetime broadening characterized by a fast tunnel rate to the reservoir  $\Gamma_\text{d}$~\cite{Ahmed2018b} and (iii) charge noise broadening, characterised by an integrated charge noise figure, $\sigma_0$, that can be dependent on the position of the QD within the channel~\cite{Kranz2020,Spence2022} (see inset in Fig.~\ref{fig:2}(a) and Appendix~\ref{app:lineshapes} for the associated lineshapes). We explore these mechanisms in Fig.~\ref{fig:2}(b) by varying the back-gate voltage, $V_\text{bs}$,  with the intention to change the position and shape of the QD within the channel and hence the tunnel rates.

First, in the top panel, we test hypothesis (i), electron-phonon decoupling, set by the relation $k_\text{B}T_\text{elect}>h\Gamma_\text{d}, \sigma_0$. We plot the extracted $T_\text{eff}$ as a function of $V_\text{bs}$ and observe a varying temperature for the range of back-gate voltages used. This measurement does not align with the expectation from electron-phonon decoupling: a constant $T_\text{d}$ as a function of $V_\text{bs}$. To test this hypothesis further, we use a separate thermometer, a calibrated PN diode placed on the same chip, to obtain an independent measurement of the electronic temperature (see Appendix~\ref{app:diode}). We find a base electronic temperature of $T_\text{elect}=640\pm 10$ mK, strongly indicating that electron-phonon decoupling is not the major cause for the lineshape broadening in the QD measurements. 

We move on to testing hypothesis (ii), lifetime broadening, that occurs when $h\Gamma_\text{d}>k_\text{B}T_\text{elect}, \sigma_0$. In the middle panel, we plot the maximum drain-source current at the top of the top hat current peak --- and its correspondent equivalent temperature $T_{\text{eff}, \Gamma}$, extracted considering symmetric tunnel rates. We observe a monotonic increase as a function of $V_\text{bs}$ and, for all $V_\text{bs}$ values, the equivalent temperature  is below the measured $T_\text{eff}$, a fact that enables discarding lifetime broadening as the cause of the elevated $T_\text{eff}$. 

Finally, in the bottom panel, we test hypothesis (iii), charge noise broadening, set by $\sigma_0 > k_\text{B}T_\text{elect}, h\Gamma_\text{d}$. To calculate the integrated charge noise figure, we use the relation $\sigma_0=\alpha_\text{g}\sigma_{V_0}$ where $\sigma_{V_0}$ is the standard deviation of the drain Fermi level position over the 20 measurements, corresponding to a measurement bandwidth between 0.005~Hz and 0.1~Hz. Here, we observe that $\sigma_0$ approaches a similar trend to the effective temperature in panel (a) in the range of measured back gate voltages, providing a strong indication that the Coulomb blockade oscillation is charge noise broadened. To extract an effective temperature from charge noise processes, we take into account that individual time traces are taken over a measurement bandwidth of 0.1 to 50 Hz and consider the common charge noise spectrum characterised by a $1/f$ dependence (see Appendix~\ref{app:noise}). We plot the associated lineshapes in  Fig.~\ref{fig:2}(c). First, we show a thermally broadened lineshape for $T=0.64$~K (blue) and a lifetime broadened lineshape for 100~pA (green). Additionally, we plot the charge noise broadened transition (orange) and an overlapping thermally broadened transition for $T=1.3$~K (blue dashed line) providing further evidence that the cause of the excess linewidth is linked to charge fluctuations. The remaining difference with $T_\text{eff}$, could be linked to a deviation from a strict $1/f$ charge noise dependence.

\section{Results}
\subsection{Static Power dissipation}

\begin{figure}
\includegraphics{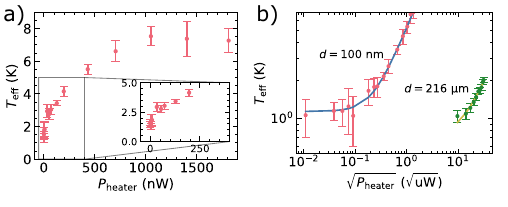}
\caption{\label{fig:3} Power budget at a distance. \textbf{(a)} Dependence of the reservoir effective temperature with power dissipation in a heater element situated at $d = \unit[100]{nm}$. (Inset) Zoom-in to the low power region. \textbf{(b)} Comparison with power dissipated at $d=\unit[216]{\mu m}$. The data is fitted using Eq.~\ref{eq:power} from which we extract the power budget before heating ($P_0$) of $\unit[45]{nW}$ and $\unit[156]{\mu W}$ respectively.}
\end{figure}

Having calibrated the QD thermometer and determined its range of operation, we now utilize the sensor to quantify the effect of local power dissipation. First, we assess the effect of static power by driving a constant drain-source current through the heating FET when $V_\text{gs}^\text{H}=0.8$~V, well above threshold. In Fig.~\ref{fig:3}(a), we plot the measured temperature as a function of static power dissipated at the heater FET for an edge-to-edge separation of 100~nm. We observe a sub linear increase in the local temperature that can be well described by a $T\propto \sqrt{P_\text{heater}}$ dependence. Such dependence is a consequence of the the quadratic dependence of the cooling power of a dilution refrigerator with temperature~\cite{Richardson1988} as well as the linear dependency of the thermal conductivity of the metals ($\propto T$) ~\cite{Duthil2014}. In Fig.~\ref{fig:3}(b), we explore this dependence in more detail as well as compare it with another measurement where the heating FET is 216~$\mu$m away. Both measurements can be well fit by the following expression, 

\begin{equation}
    T = \beta \sqrt{P_{\mathrm{heater}}+P_0},
    \label{eq:power}
\end{equation}

\noindent where $\beta$ is a temperature-power proportionality factor and $P_0$ is the dissipated power that increases the local temperature by a factor $\sqrt{2}$. The parameter $P_0$ can be then used as a proxy for the power dissipation that starts to significantly affect the operation of a QD device. We obtain $P_0=45\pm5$~nW for $d=100$~nm and $P_0=157\pm21~\mu$W for $d=216$~$\mu$m. The results indicate that, at close distances of the order of 100~nm, the power budget is limited but sufficient to enable co-integration of simple digital circuits like multiplexers~\cite{Acharya2023}. For larger distances, of the order of 100 ~$\mu$m, still well within the dimensions of common chip sizes, the power budget should be suitable to accommodate some complex cryoCMOS elements like transimpedance amplifiers~\cite{LeGuevel2020} and active inductors~\cite{LEGUEVEL2023} but not the most advanced controllers~\cite{Xue2021} and receivers \cite{Ruffino2021}. Additionally, we extract a $\beta=(8\pm2)\cdot 10^3$~KW$^{-1/2}$ for $d=100$~nm and $\beta=(59\pm1)$~KW$^{-1/2}$ for $d=216$~$\mu$m. As expected, the temperature-power proportionally factor $\beta$ is larger the closer the heating element is. 

\subsection{Dynamic Power Dissipation}

\begin{figure}
\includegraphics{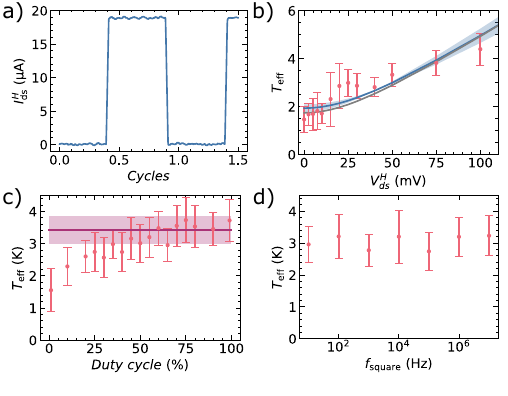}
\caption{\label{fig:4} Dynamic power dissipation. In \textbf{(a)}, we show the current through the heater element for a \unit[1]{kHz} square wave applied to the drain ohmic of the heater. \textbf{(b)} Shows the effective reservoir temperature for different heater source pulse depths, measured at \unit[1]{kHz} and at $d$ = \unit[100]{nm}. Fit to Eq.~\ref{eq:power} in blue. \textbf{(c)} Effective temperature vs duty cycle for $f_\text{square}$= 1~kHz and $V_\text{ds}^\text{H}=25$~mV. The red line and stripe indicates the static power dissipation and standard deviation at the same $V_\text{ds}^\text{H}$ amplitude. \textbf{(d)} Effective temperature vs the square wave frequency showing a flat response.}
\end{figure}

Although continuous drive currents, like those used here to produce varying levels of static dissipated power, are common in analog circuit applications, they are uncommon in digital circuits. In such circuits, dynamic power dissipation could be the dominant contributor. In this Section, we focus on quantifying the effects of dynamical power dissipation for switching rates of the heating FET up to 10~MHz. For this experiment, we focus on the $d=100$~nm sample. We send square wave voltage pulses to the source ohmic of the heater element and measure the resulting current through an in-line resistor, see Fig.~\ref{fig:4}(a). The pulses are characterised by a voltage amplitude $V_\text{ds}^\text{H}$, frequency $f_\text{square}$, and duty cycle $D$. We first vary the voltage pulse amplitude $V_\text{ds}^\text{H}<0.1$~V for $f_\text{square}=1$~kHz and $D=50$\%. 

First, we look at the dependence with voltage amplitude in Fig.~\ref{fig:4}(b). We observe an increase in temperature that becomes linear for pulse amplitudes above 20 mV. The trend matches well the expectation set by Eq.~\ref{eq:power}, given that the heating FET is operated in the linear regime.
We include a fit to the data and the associated standard deviation in the shaded region. The gray fit is obtained by fitting the same equation but excluding the 4 points between 15 and 30 mV which visually look to be outliers. Next, in Fig~\ref{fig:4}(c), we change the duty cycle --- a higher duty cycle indicates a longer on-state time for the heater--- for a fixed $f_\text{square}=1$~kHz and $V_\text{ds}^\text{H}=25$~mV. We observe how the temperature increases until it overlaps with the value produced with static dissipated power at the given $V_\text{ds}^\text{H}$, indicated by the red horizontal line. Finally, in Fig~\ref{fig:4}(d), we record the temperature as we increase $f_\text{square}$ up to 10~MHz at a constant duty cycle $D=50$\% and voltage amplitude $V_\text{sd}^\text{H}=25$~mV. We observe no strong frequency dependence in the extracted temperature. Given the transistor dimensions and the pulse amplitude, we estimate an upper bound for the dissipated dynamic power of 1~pW. This result indicates that, within the measurement resolution, such dynamic power dissipation level has a negligible contribution to the total power dissipated in the system. Hence, it is expected that at higher frequencies, larger number of switching devices and pulse amplitudes will be tolerable up to the tens of nanowatts as observed in the static power dissipation measurements. For example, operating 1000 FETs at 10~GHz using the same pulsing conditions as described above could be manageable.   

\section{Conclusions and Outlook}

We have demonstrated a method to characterize the impact of active transistor elements on local temperature using a QD-based thermometer. We found that 45~nW is enough to elevate the local temperature at a distance of 100~nm whereas the power budget can be increased to $\approx150$~$\mu$W at 216~$\mu$m. Such double FET structures, manufactured using an industrial FDSOI process, may enable further studies of the mechanisms involved in thermal transport by exploring in more detail the dependence with device separation. Furthermore, such QD thermometers, when measured using high-bandwidth reflectometry techniques~\cite{vigneau2022} may enable the study of time-dependent thermal phenomena as well as quantum thermodynamics experiments~\cite{champain2023realtime}. \\

\section{Acknowledgements}
Quantum Motion would like to thank Nigel Cave at GlobalFoundries for helpful discussions. All authors acknowledge Jonathan Warren and James Kirkman of Quantum Motion for their technical support during this work. M.F.G.Z. acknowledges a UKRI Future Leaders Fellowship [MR/V023284/1]. 

\appendix
 
\section{Methods:}
\label{app:methods}
The experiments in this work are performed in a cryogen-free dilution refrigerator with a base temperature of 20~mK. The individual devices are accessed through an integrated on-chip multiplexer (MUX) with an equivalent on-state series resistance below 10~k$\Omega$. RC low-pass filters with a cutoff frequency of 65~kHz are installed at the mixing chamber (MXC), for square wave pulsing on the heater FET these filters are bypassed. At room temperature, two current-to-voltage amplifiers (BasPI SP983c) are used to amplify the current from the quantum dot (QD) channel subsequently a spectrum Mi4 ADC card is used to digitize the signals. A QDAC-II by QDevil is used as a stable precision voltage source to supply voltages to the gates.

\section{Impact of excited states}
\label{app:Impact of excited states}

In this section,  we describe the impact of excited states in the measurement bias window, i.e. when $eV_\text{ds}> \Delta E^n$, where $\Delta E^n$ is the difference in energy between the ground and n$^{th}$ excited state. In this case, in the limit of fast intradot relaxation rates, the drain-source current expression needs to be modified to include the additional electrical transport channels into the QD due to the additional discrete states i.e. the regime of  sequential multi-level transport. The current expression then reads,

\begin{equation}
    I_{\mathrm{ds}} = e\frac{\Gamma_{\mathrm{out}}\cdot\sum_n \Gamma_{\mathrm{in,n}}}{\Gamma_{\mathrm{out}}+\sum_n \Gamma_{\mathrm{in,n}}}.
    \label{eq:exc_states}
\end{equation}

\begin{figure}
    \centering
    \includegraphics[width=1\linewidth]{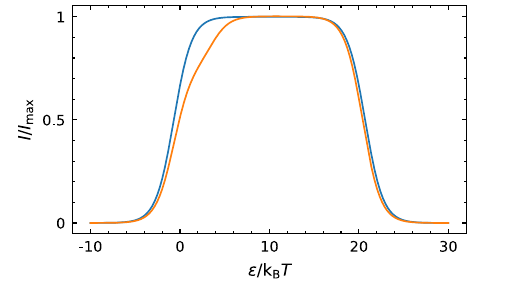}
    \caption{Simulation of the effect of excited states in the bias window. (Blue) Top hat lineshape simulated using Eq.~\ref{eq:tophat} using $\Gamma_\text{in,0}=\Gamma_\text{out,0}$ and $\alpha_\text{g}[V_{0,\text{out}}-V_{0,\text{in}}]=10$. (Orange) Top hat lineshape simulated using Eq.~\ref{eq:exc_states} adding $\Gamma_\text{in,1}=1$ and $\alpha_\text{g}[V_{0,\text{out,1}}-V_{0,\text{in,0}}]=2$ .  }
    \label{fig:excited}
\end{figure}

As we can see in Fig.~\ref{fig:excited} the rising edge of the IV curve presents a staircase behaviour that may affect the accurate extraction of the relevant temperature. However, the falling edge remains unaffected and hence the temperature can be extracted accurately.  

\section{Electronic temperature}
\label{app:diode}

In this section, we present an independent measurement of the electronic temperature performed using a calibrated on-chip PN diode. The calibration is done with respect to the MXC thermometer reading, assuming perfect thermalization. However, the transistors used in this study are electrically accessed through an on-chip cryogenic multiplexer that requires powering up support digital and analogue electronics. The power consumed by the support circuitry when measuring the FETs is 4.3~$\mu$W and elevates the chip temperature above that of the MXC. 

To measure the on-chip temperature, we bias the diode with a 1~nA bias current and measure the voltage drop across it, which has been previously calibrated to the MXC temperature when the supporting electronics are off. We fit the data to the expression,

\begin{equation}
    T_\text{Diode}= [(T_\text{MXC}+T_1)^n+T_0^n]^{1/n},
    \label{eq:diode}
\end{equation}

where $T_0$ is the minimum on-chip temperature when the support electronics are on and $T_1$ is the on-chip temperature elevation above $T_\text{MXC}$ at high temperatures. A fit to the data reveals a $T_0=640\pm10$~mK,  $T_1=258\pm11$ mK and $n=11.0\pm 0.2$. The lower electronic temperature compared to the values extracted for the QD effective temperature in Fig.\ref{fig:2}(c), reemphasize the idea that mechanisms other than electron-phonon decoupling set the lineshape of the Coulomb blockade oscillations.  
\begin{figure}
\includegraphics{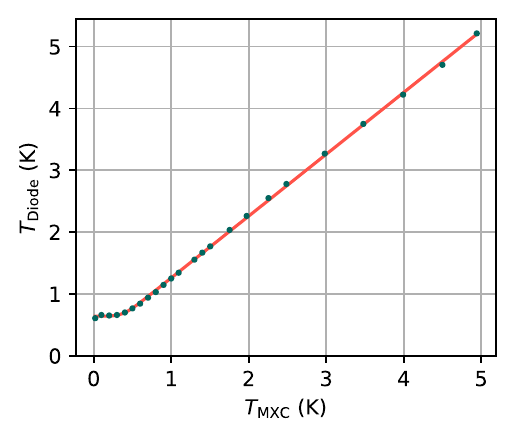}
\caption{\label{fig:6}Diode thermometry.  Diode temperature versus MXC temperature (green dots) and fit (red line). }
\end{figure}

\section{Lineshapes in the different transport regimes}
\label{app:lineshapes}

Here, we describe the different lineshapes depending on specific energy scale dominating the electrical transport through a discrete state in the QD. To facilitate the visual comparison, we focus on the differences in the tunnel rate expression since, close to the degeneracy point between the electrochemical potential of the QD and one of the reservoirs, the current reads $I_\text{ds}\approx e\Gamma$.   

\textbf{Thermal broadening regime ($k_\text{B}T>h\Gamma, \sigma_0$) }: In this case, the tunnel rate expression tracks the number of states in the reservoir which is proportional to the Fermi function in the wide-band limit where the denisty of state is weakly dependent on energy detuning ($\varepsilon$),  

\begin{equation}
    \Gamma_\text{Thermal}=\frac{\Gamma_0}{1+\exp(-\varepsilon/k_\text{B}T)},
    \label{eq:thermal}
\end{equation}

\noindent where $\Gamma_0$ is the maximum tunnel rate occurring at large positive detuning. 

\textbf{Lifetime broadening regime  ($h\Gamma> k_\text{B}T, \sigma_0$) }: In this case, the tunnel rate expression is modified by the finite width of the discrete state of the QD due to Heisenberg's uncertainty principle,

\begin{equation}
    \Gamma_\text{Lifetime}=\frac{\Gamma_0}{\pi}[\arctan(\varepsilon/\Gamma_0)+\pi/2].
    \label{eq:lifetime}
\end{equation}

In principle, the two mechanisms (thermal and lifetime) can coexist, in which case the tunnel rate is caused by a convolution of the two processes $\Gamma=\Gamma_\text{Thermal}*\Gamma_\text{Lifetime}$~\cite{Chawner2021, peri2023beyondadiabatic} . 

\textbf{Charge noise broadening regime  ($\sigma_0> k_\text{B}T, h\Gamma$):} Charge noise has the effect of introducing fluctuations on the detuning between the QD electrochemical level and the charge reservoir. Effectively, this corresponds to sampling from a Gaussian distribution of detuning values with zero expectation value and corresponding standard deviation of $\sigma_0$, the integrated charge noise over the bandwidth of the measurement~\cite{Kranz2020}. The resulting tunnel rate (and IV lineshape) can be obtained by the convolution of the relevant tunnel rate expression and a Gaussian, $g(\mu=0$, $\sigma^2=\sigma_0^2)$,

\begin{equation}
    \Gamma_\text{Charge noise}=\Gamma_\text{Thermal/Lifetime}*g. 
    \label{eq:convolution}
\end{equation}

\section{Charge noise analysis}
\label{app:noise}

In Fig.~\ref{fig:2}(c), we calculated the equivalent temperature for a charge noise process characterised by an integrated charge noise figure over the bandwidth 0.005 to 0.1 Hz of $\sigma_0=\alpha_\text{g}\cdot 100$~$\mu$V. To perform the extraction, we first consider the impact of $1/f$ charge noise over the individual top hat traces which are acquired over a bandwidth of 0.1 to 50~Hz. We estimate that over this bandwidth, the integrated charge noise figure corresponds to $\sigma_{0,H}=\alpha_\text{g}\cdot 150$~$\mu$V. By using  Eq.~\ref{eq:convolution}, we find an equivalent temperature, $T=1.12$~K, for the individual Coulomb blockade traces. Next, we simulate the effect of the measurement over the 20 top hat traces and find that the final equivalent temperature corresponds to $T=1.3$~K. This result approaches the experimental observation in Fig~\ref{fig:2}(b) (top panel). The remaining differences could be associated to a discrepancy with the assumed $1/f$ charge noise spectrum. 

\bibliography{biblio}

\begin{thebibliography}{47}%
\makeatletter
\providecommand \@ifxundefined [1]{%
 \@ifx{#1\undefined}
}%
\providecommand \@ifnum [1]{%
 \ifnum #1\expandafter \@firstoftwo
 \else \expandafter \@secondoftwo
 \fi
}%
\providecommand \@ifx [1]{%
 \ifx #1\expandafter \@firstoftwo
 \else \expandafter \@secondoftwo
 \fi
}%
\providecommand \natexlab [1]{#1}%
\providecommand \enquote  [1]{``#1''}%
\providecommand \bibnamefont  [1]{#1}%
\providecommand \bibfnamefont [1]{#1}%
\providecommand \citenamefont [1]{#1}%
\providecommand \href@noop [0]{\@secondoftwo}%
\providecommand \href [0]{\begingroup \@sanitize@url \@href}%
\providecommand \@href[1]{\@@startlink{#1}\@@href}%
\providecommand \@@href[1]{\endgroup#1\@@endlink}%
\providecommand \@sanitize@url [0]{\catcode `\\12\catcode `\$12\catcode `\&12\catcode `\#12\catcode `\^12\catcode `\_12\catcode `\%12\relax}%
\providecommand \@@startlink[1]{}%
\providecommand \@@endlink[0]{}%
\providecommand \url  [0]{\begingroup\@sanitize@url \@url }%
\providecommand \@url [1]{\endgroup\@href {#1}{\urlprefix }}%
\providecommand \urlprefix  [0]{URL }%
\providecommand \Eprint [0]{\href }%
\providecommand \doibase [0]{https://doi.org/}%
\providecommand \selectlanguage [0]{\@gobble}%
\providecommand \bibinfo  [0]{\@secondoftwo}%
\providecommand \bibfield  [0]{\@secondoftwo}%
\providecommand \translation [1]{[#1]}%
\providecommand \BibitemOpen [0]{}%
\providecommand \bibitemStop [0]{}%
\providecommand \bibitemNoStop [0]{.\EOS\space}%
\providecommand \EOS [0]{\spacefactor3000\relax}%
\providecommand \BibitemShut  [1]{\csname bibitem#1\endcsname}%
\let\auto@bib@innerbib\@empty
\bibitem [{\citenamefont {Gonzalez-Zalba}\ \emph {et~al.}(2021)\citenamefont {Gonzalez-Zalba}, \citenamefont {de~Franceschi}, \citenamefont {Charbon}, \citenamefont {Meunier}, \citenamefont {Vinet},\ and\ \citenamefont {Dzurak}}]{GonzalezZalba2021}%
  \BibitemOpen
  \bibfield  {author} {\bibinfo {author} {\bibfnamefont {M.~F.}\ \bibnamefont {Gonzalez-Zalba}}, \bibinfo {author} {\bibfnamefont {S.}~\bibnamefont {de~Franceschi}}, \bibinfo {author} {\bibfnamefont {E.}~\bibnamefont {Charbon}}, \bibinfo {author} {\bibfnamefont {T.}~\bibnamefont {Meunier}}, \bibinfo {author} {\bibfnamefont {M.}~\bibnamefont {Vinet}},\ and\ \bibinfo {author} {\bibfnamefont {A.~S.}\ \bibnamefont {Dzurak}},\ }\bibfield  {title} {\bibinfo {title} {Scaling silicon-based quantum computing using cmos technology},\ }\href@noop {} {\bibfield  {journal} {\bibinfo  {journal} {Nat. Electron.}\ }\textbf {\bibinfo {volume} {4}},\ \bibinfo {pages} {872} (\bibinfo {year} {2021})}\BibitemShut {NoStop}%
\bibitem [{\citenamefont {Xue}\ \emph {et~al.}(2022)\citenamefont {Xue}, \citenamefont {Russ}, \citenamefont {Samkharadze}, \citenamefont {Undseth}, \citenamefont {Sammak}, \citenamefont {Scappucci},\ and\ \citenamefont {Vandersypen}}]{Xue2022}%
  \BibitemOpen
  \bibfield  {author} {\bibinfo {author} {\bibfnamefont {X.}~\bibnamefont {Xue}}, \bibinfo {author} {\bibfnamefont {M.}~\bibnamefont {Russ}}, \bibinfo {author} {\bibfnamefont {N.}~\bibnamefont {Samkharadze}}, \bibinfo {author} {\bibfnamefont {B.}~\bibnamefont {Undseth}}, \bibinfo {author} {\bibfnamefont {A.}~\bibnamefont {Sammak}}, \bibinfo {author} {\bibfnamefont {G.}~\bibnamefont {Scappucci}},\ and\ \bibinfo {author} {\bibfnamefont {L.~M.~K.}\ \bibnamefont {Vandersypen}},\ }\bibfield  {title} {\bibinfo {title} {Quantum logic with spin qubits crossing the surface code threshold},\ }\href@noop {} {\bibfield  {journal} {\bibinfo  {journal} {Nature}\ }\textbf {\bibinfo {volume} {601}},\ \bibinfo {pages} {343} (\bibinfo {year} {2022})}\BibitemShut {NoStop}%
\bibitem [{\citenamefont {Noiri}\ \emph {et~al.}(2022)\citenamefont {Noiri}, \citenamefont {Takeda}, \citenamefont {Nakajima}, \citenamefont {Kobayashi}, \citenamefont {Sammak}, \citenamefont {Scappucci},\ and\ \citenamefont {Tarucha}}]{Noiri2022}%
  \BibitemOpen
  \bibfield  {author} {\bibinfo {author} {\bibfnamefont {A.}~\bibnamefont {Noiri}}, \bibinfo {author} {\bibfnamefont {K.}~\bibnamefont {Takeda}}, \bibinfo {author} {\bibfnamefont {T.}~\bibnamefont {Nakajima}}, \bibinfo {author} {\bibfnamefont {T.}~\bibnamefont {Kobayashi}}, \bibinfo {author} {\bibfnamefont {A.}~\bibnamefont {Sammak}}, \bibinfo {author} {\bibfnamefont {G.}~\bibnamefont {Scappucci}},\ and\ \bibinfo {author} {\bibfnamefont {S.}~\bibnamefont {Tarucha}},\ }\bibfield  {title} {\bibinfo {title} {Fast universal quantum gate above the fault-tolerance threshold in silicon},\ }\href@noop {} {\bibfield  {journal} {\bibinfo  {journal} {Nature}\ }\textbf {\bibinfo {volume} {601}},\ \bibinfo {pages} {338} (\bibinfo {year} {2022})}\BibitemShut {NoStop}%
\bibitem [{\citenamefont {Mills}\ \emph {et~al.}(2022)\citenamefont {Mills}, \citenamefont {Guinn}, \citenamefont {Gullans}, \citenamefont {Sigillito}, \citenamefont {Feldman}, \citenamefont {Nielsen},\ and\ \citenamefont {Petta}}]{Mills2022}%
  \BibitemOpen
  \bibfield  {author} {\bibinfo {author} {\bibfnamefont {A.~R.}\ \bibnamefont {Mills}}, \bibinfo {author} {\bibfnamefont {C.~R.}\ \bibnamefont {Guinn}}, \bibinfo {author} {\bibfnamefont {M.~J.}\ \bibnamefont {Gullans}}, \bibinfo {author} {\bibfnamefont {A.~J.}\ \bibnamefont {Sigillito}}, \bibinfo {author} {\bibfnamefont {M.~M.}\ \bibnamefont {Feldman}}, \bibinfo {author} {\bibfnamefont {E.}~\bibnamefont {Nielsen}},\ and\ \bibinfo {author} {\bibfnamefont {J.~R.}\ \bibnamefont {Petta}},\ }\bibfield  {title} {\bibinfo {title} {Two-qubit silicon quantum processor with operation fidelity exceeding 99\%},\ }\href {https://doi.org/10.1126/sciadv.abn5130} {\bibfield  {journal} {\bibinfo  {journal} {Science Advances}\ }\textbf {\bibinfo {volume} {8}},\ \bibinfo {pages} {eabn5130} (\bibinfo {year} {2022})}\BibitemShut {NoStop}%
\bibitem [{\citenamefont {Takeda}\ \emph {et~al.}(2022)\citenamefont {Takeda}, \citenamefont {Noiri}, \citenamefont {Nakajima}, \citenamefont {Kobayashi},\ and\ \citenamefont {Tarucha}}]{Takeda2022}%
  \BibitemOpen
  \bibfield  {author} {\bibinfo {author} {\bibfnamefont {K.}~\bibnamefont {Takeda}}, \bibinfo {author} {\bibfnamefont {A.}~\bibnamefont {Noiri}}, \bibinfo {author} {\bibfnamefont {T.}~\bibnamefont {Nakajima}}, \bibinfo {author} {\bibfnamefont {T.}~\bibnamefont {Kobayashi}},\ and\ \bibinfo {author} {\bibfnamefont {S.}~\bibnamefont {Tarucha}},\ }\bibfield  {title} {\bibinfo {title} {Quantum error correction with silicon spin qubits},\ }\href {https://doi.org/10.1038/s41586-022-04986-6} {\bibfield  {journal} {\bibinfo  {journal} {Nature}\ }\textbf {\bibinfo {volume} {608}},\ \bibinfo {pages} {682} (\bibinfo {year} {2022})}\BibitemShut {NoStop}%
\bibitem [{\citenamefont {Philips}\ \emph {et~al.}(2022)\citenamefont {Philips}, \citenamefont {M{\k{a}}dzik}, \citenamefont {Amitonov}, \citenamefont {de~Snoo}, \citenamefont {Russ}, \citenamefont {Kalhor}, \citenamefont {Volk}, \citenamefont {Lawrie}, \citenamefont {Brousse}, \citenamefont {Tryputen}, \citenamefont {Wuetz}, \citenamefont {Sammak}, \citenamefont {Veldhorst}, \citenamefont {Scappucci},\ and\ \citenamefont {Vandersypen}}]{Philips2022}%
  \BibitemOpen
  \bibfield  {author} {\bibinfo {author} {\bibfnamefont {S.~G.~J.}\ \bibnamefont {Philips}}, \bibinfo {author} {\bibfnamefont {M.~T.}\ \bibnamefont {M{\k{a}}dzik}}, \bibinfo {author} {\bibfnamefont {S.~V.}\ \bibnamefont {Amitonov}}, \bibinfo {author} {\bibfnamefont {S.~L.}\ \bibnamefont {de~Snoo}}, \bibinfo {author} {\bibfnamefont {M.}~\bibnamefont {Russ}}, \bibinfo {author} {\bibfnamefont {N.}~\bibnamefont {Kalhor}}, \bibinfo {author} {\bibfnamefont {C.}~\bibnamefont {Volk}}, \bibinfo {author} {\bibfnamefont {W.~I.~L.}\ \bibnamefont {Lawrie}}, \bibinfo {author} {\bibfnamefont {D.}~\bibnamefont {Brousse}}, \bibinfo {author} {\bibfnamefont {L.}~\bibnamefont {Tryputen}}, \bibinfo {author} {\bibfnamefont {B.~P.}\ \bibnamefont {Wuetz}}, \bibinfo {author} {\bibfnamefont {A.}~\bibnamefont {Sammak}}, \bibinfo {author} {\bibfnamefont {M.}~\bibnamefont {Veldhorst}}, \bibinfo {author} {\bibfnamefont {G.}~\bibnamefont {Scappucci}},\ and\ \bibinfo {author} {\bibfnamefont {L.~M.~K.}\ \bibnamefont {Vandersypen}},\
  }\bibfield  {title} {\bibinfo {title} {Universal control of a six-qubit quantum processor in silicon},\ }\href {https://doi.org/10.1038/s41586-022-05117-x} {\bibfield  {journal} {\bibinfo  {journal} {Nature}\ }\textbf {\bibinfo {volume} {609}},\ \bibinfo {pages} {919} (\bibinfo {year} {2022})}\BibitemShut {NoStop}%
\bibitem [{\citenamefont {Maurand}\ \emph {et~al.}(2016)\citenamefont {Maurand}, \citenamefont {Jehl}, \citenamefont {Kotekar-Patil}, \citenamefont {Corna}, \citenamefont {Bohuslavskyi}, \citenamefont {Lavi{\'e}ville}, \citenamefont {Hutin}, \citenamefont {Barraud}, \citenamefont {Vinet}, \citenamefont {Sanquer},\ and\ \citenamefont {De~Franceschi}}]{Maurand2016}%
  \BibitemOpen
  \bibfield  {author} {\bibinfo {author} {\bibfnamefont {R.}~\bibnamefont {Maurand}}, \bibinfo {author} {\bibfnamefont {X.}~\bibnamefont {Jehl}}, \bibinfo {author} {\bibfnamefont {D.}~\bibnamefont {Kotekar-Patil}}, \bibinfo {author} {\bibfnamefont {A.}~\bibnamefont {Corna}}, \bibinfo {author} {\bibfnamefont {H.}~\bibnamefont {Bohuslavskyi}}, \bibinfo {author} {\bibfnamefont {R.}~\bibnamefont {Lavi{\'e}ville}}, \bibinfo {author} {\bibfnamefont {L.}~\bibnamefont {Hutin}}, \bibinfo {author} {\bibfnamefont {S.}~\bibnamefont {Barraud}}, \bibinfo {author} {\bibfnamefont {M.}~\bibnamefont {Vinet}}, \bibinfo {author} {\bibfnamefont {M.}~\bibnamefont {Sanquer}},\ and\ \bibinfo {author} {\bibfnamefont {S.}~\bibnamefont {De~Franceschi}},\ }\bibfield  {title} {\bibinfo {title} {A cmos silicon spin qubit},\ }\href {http://dx.doi.org/10.1038/ncomms13575} {\bibfield  {journal} {\bibinfo  {journal} {Nature Communications}\ }\textbf {\bibinfo {volume} {7}},\ \bibinfo {pages} {13575 EP } (\bibinfo {year} {2016})},\ \bibinfo
  {note} {article}\BibitemShut {NoStop}%
\bibitem [{\citenamefont {Zwerver}\ \emph {et~al.}(2022)\citenamefont {Zwerver}, \citenamefont {Krahenmann}, \citenamefont {Watson}, \citenamefont {Lampert}, \citenamefont {George}, \citenamefont {Pillarisetty}, \citenamefont {Bojarski}, \citenamefont {Amin}, \citenamefont {Amitonov}, \citenamefont {Boter}, \citenamefont {Caudillo}, \citenamefont {Correas-Serrano}, \citenamefont {Dehollain}, \citenamefont {Droulers}, \citenamefont {Henry}, \citenamefont {Kotlyar}, \citenamefont {Lodari}, \citenamefont {L{\"u}thi}, \citenamefont {Michalak}, \citenamefont {Mueller}, \citenamefont {Neyens}, \citenamefont {Roberts}, \citenamefont {Samkharadze}, \citenamefont {Zheng}, \citenamefont {Zietz}, \citenamefont {Scappucci}, \citenamefont {Veldhorst}, \citenamefont {Vandersypen},\ and\ \citenamefont {Clarke}}]{Zwerver2022}%
  \BibitemOpen
  \bibfield  {author} {\bibinfo {author} {\bibfnamefont {A.~M.~J.}\ \bibnamefont {Zwerver}}, \bibinfo {author} {\bibfnamefont {T.}~\bibnamefont {Krahenmann}}, \bibinfo {author} {\bibfnamefont {T.~F.}\ \bibnamefont {Watson}}, \bibinfo {author} {\bibfnamefont {L.}~\bibnamefont {Lampert}}, \bibinfo {author} {\bibfnamefont {H.~C.}\ \bibnamefont {George}}, \bibinfo {author} {\bibfnamefont {R.}~\bibnamefont {Pillarisetty}}, \bibinfo {author} {\bibfnamefont {S.~A.}\ \bibnamefont {Bojarski}}, \bibinfo {author} {\bibfnamefont {P.}~\bibnamefont {Amin}}, \bibinfo {author} {\bibfnamefont {S.~V.}\ \bibnamefont {Amitonov}}, \bibinfo {author} {\bibfnamefont {J.~M.}\ \bibnamefont {Boter}}, \bibinfo {author} {\bibfnamefont {R.}~\bibnamefont {Caudillo}}, \bibinfo {author} {\bibfnamefont {D.}~\bibnamefont {Correas-Serrano}}, \bibinfo {author} {\bibfnamefont {J.~P.}\ \bibnamefont {Dehollain}}, \bibinfo {author} {\bibfnamefont {G.}~\bibnamefont {Droulers}}, \bibinfo {author} {\bibfnamefont {E.~M.}\ \bibnamefont {Henry}}, \bibinfo
  {author} {\bibfnamefont {R.}~\bibnamefont {Kotlyar}}, \bibinfo {author} {\bibfnamefont {M.}~\bibnamefont {Lodari}}, \bibinfo {author} {\bibfnamefont {F.}~\bibnamefont {L{\"u}thi}}, \bibinfo {author} {\bibfnamefont {D.~J.}\ \bibnamefont {Michalak}}, \bibinfo {author} {\bibfnamefont {B.~K.}\ \bibnamefont {Mueller}}, \bibinfo {author} {\bibfnamefont {S.}~\bibnamefont {Neyens}}, \bibinfo {author} {\bibfnamefont {J.}~\bibnamefont {Roberts}}, \bibinfo {author} {\bibfnamefont {N.}~\bibnamefont {Samkharadze}}, \bibinfo {author} {\bibfnamefont {G.}~\bibnamefont {Zheng}}, \bibinfo {author} {\bibfnamefont {O.~K.}\ \bibnamefont {Zietz}}, \bibinfo {author} {\bibfnamefont {G.}~\bibnamefont {Scappucci}}, \bibinfo {author} {\bibfnamefont {M.}~\bibnamefont {Veldhorst}}, \bibinfo {author} {\bibfnamefont {L.~M.~K.}\ \bibnamefont {Vandersypen}},\ and\ \bibinfo {author} {\bibfnamefont {J.~S.}\ \bibnamefont {Clarke}},\ }\bibfield  {title} {\bibinfo {title} {Qubits made by advanced semiconductor manufacturing},\ }\href
  {https://doi.org/10.1038/s41928-022-00727-9} {\bibfield  {journal} {\bibinfo  {journal} {Nature Electronics}\ }\textbf {\bibinfo {volume} {5}},\ \bibinfo {pages} {184} (\bibinfo {year} {2022})}\BibitemShut {NoStop}%
\bibitem [{\citenamefont {Camenzind}\ \emph {et~al.}(2022)\citenamefont {Camenzind}, \citenamefont {Geyer}, \citenamefont {Fuhrer}, \citenamefont {Warburton}, \citenamefont {Zumb{\"u}hl},\ and\ \citenamefont {Kuhlmann}}]{Camenzind2021}%
  \BibitemOpen
  \bibfield  {author} {\bibinfo {author} {\bibfnamefont {L.~C.}\ \bibnamefont {Camenzind}}, \bibinfo {author} {\bibfnamefont {S.}~\bibnamefont {Geyer}}, \bibinfo {author} {\bibfnamefont {A.}~\bibnamefont {Fuhrer}}, \bibinfo {author} {\bibfnamefont {R.~J.}\ \bibnamefont {Warburton}}, \bibinfo {author} {\bibfnamefont {D.~M.}\ \bibnamefont {Zumb{\"u}hl}},\ and\ \bibinfo {author} {\bibfnamefont {A.~V.}\ \bibnamefont {Kuhlmann}},\ }\bibfield  {title} {\bibinfo {title} {A hole spin qubit in a fin field-effect transistor above 4 kelvin},\ }\href@noop {} {\bibfield  {journal} {\bibinfo  {journal} {Nature Electronics}\ }\textbf {\bibinfo {volume} {5}},\ \bibinfo {pages} {178} (\bibinfo {year} {2022})}\BibitemShut {NoStop}%
\bibitem [{\citenamefont {Schaal}\ \emph {et~al.}(2019)\citenamefont {Schaal}, \citenamefont {Rossi}, \citenamefont {Ciriano-Tejel}, \citenamefont {Yang}, \citenamefont {Barraud}, \citenamefont {Morton},\ and\ \citenamefont {Gonzalez-Zalba}}]{Schaal2019}%
  \BibitemOpen
  \bibfield  {author} {\bibinfo {author} {\bibfnamefont {S.}~\bibnamefont {Schaal}}, \bibinfo {author} {\bibfnamefont {A.}~\bibnamefont {Rossi}}, \bibinfo {author} {\bibfnamefont {V.}~\bibnamefont {Ciriano-Tejel}}, \bibinfo {author} {\bibfnamefont {T.-Y.}\ \bibnamefont {Yang}}, \bibinfo {author} {\bibfnamefont {S.}~\bibnamefont {Barraud}}, \bibinfo {author} {\bibfnamefont {J.}~\bibnamefont {Morton}},\ and\ \bibinfo {author} {\bibfnamefont {M.}~\bibnamefont {Gonzalez-Zalba}},\ }\bibfield  {title} {\bibinfo {title} {{A CMOS dynamic random access architecture for radio-frequency readout of quantum devices}},\ }\bibfield  {journal} {\bibinfo  {journal} {Nature Electronics}\ }\textbf {\bibinfo {volume} {2}},\ \href {https://doi.org/10.1038/s41928-019-0259-5} {10.1038/s41928-019-0259-5} (\bibinfo {year} {2019})\BibitemShut {NoStop}%
\bibitem [{\citenamefont {Ruffino}\ \emph {et~al.}(2022)\citenamefont {Ruffino}, \citenamefont {Yang}, \citenamefont {Michniewicz}, \citenamefont {Peng}, \citenamefont {Charbon},\ and\ \citenamefont {Gonzalez-Zalba}}]{Ruffino2022}%
  \BibitemOpen
  \bibfield  {author} {\bibinfo {author} {\bibfnamefont {A.}~\bibnamefont {Ruffino}}, \bibinfo {author} {\bibfnamefont {T.-Y.}\ \bibnamefont {Yang}}, \bibinfo {author} {\bibfnamefont {J.}~\bibnamefont {Michniewicz}}, \bibinfo {author} {\bibfnamefont {Y.}~\bibnamefont {Peng}}, \bibinfo {author} {\bibfnamefont {E.}~\bibnamefont {Charbon}},\ and\ \bibinfo {author} {\bibfnamefont {M.~F.}\ \bibnamefont {Gonzalez-Zalba}},\ }\bibfield  {title} {\bibinfo {title} {A cryo-cmos chip that integrates silicon quantum dots and multiplexed dispersive readout electronics},\ }\href@noop {} {\bibfield  {journal} {\bibinfo  {journal} {Nat. Electron.}\ }\textbf {\bibinfo {volume} {5}},\ \bibinfo {pages} {53} (\bibinfo {year} {2022})}\BibitemShut {NoStop}%
\bibitem [{\citenamefont {Reilly}(2019)}]{Reilly2019}%
  \BibitemOpen
  \bibfield  {author} {\bibinfo {author} {\bibfnamefont {D.~J.}\ \bibnamefont {Reilly}},\ }\bibfield  {title} {\bibinfo {title} {Challenges in scaling-up the control interface of a quantum computer},\ }in\ \href {https://doi.org/10.1109/IEDM19573.2019.8993497} {\emph {\bibinfo {booktitle} {2019 IEEE International Electron Devices Meeting (IEDM)}}}\ (\bibinfo {year} {2019})\ pp.\ \bibinfo {pages} {31.7.1--31.7.6}\BibitemShut {NoStop}%
\bibitem [{\citenamefont {van Dijk}\ \emph {et~al.}(2019)\citenamefont {van Dijk}, \citenamefont {Kawakami}, \citenamefont {Schouten}, \citenamefont {Veldhorst}, \citenamefont {Vandersypen}, \citenamefont {Babaie}, \citenamefont {Charbon},\ and\ \citenamefont {Sebastiano}}]{Dijk2019}%
  \BibitemOpen
  \bibfield  {author} {\bibinfo {author} {\bibfnamefont {J.}~\bibnamefont {van Dijk}}, \bibinfo {author} {\bibfnamefont {E.}~\bibnamefont {Kawakami}}, \bibinfo {author} {\bibfnamefont {R.}~\bibnamefont {Schouten}}, \bibinfo {author} {\bibfnamefont {M.}~\bibnamefont {Veldhorst}}, \bibinfo {author} {\bibfnamefont {L.}~\bibnamefont {Vandersypen}}, \bibinfo {author} {\bibfnamefont {M.}~\bibnamefont {Babaie}}, \bibinfo {author} {\bibfnamefont {E.}~\bibnamefont {Charbon}},\ and\ \bibinfo {author} {\bibfnamefont {F.}~\bibnamefont {Sebastiano}},\ }\bibfield  {title} {\bibinfo {title} {Impact of classical control electronics on qubit fidelity},\ }\href {https://doi.org/10.1103/physrevapplied.12.044054} {\bibfield  {journal} {\bibinfo  {journal} {Physical Review Applied}\ }\textbf {\bibinfo {volume} {12}},\ \bibinfo {pages} {044054} (\bibinfo {year} {2019})}\BibitemShut {NoStop}%
\bibitem [{\citenamefont {Xue}\ \emph {et~al.}(2021)\citenamefont {Xue}, \citenamefont {Patra}, \citenamefont {van Dijk}, \citenamefont {Samkharadze}, \citenamefont {Subramanian}, \citenamefont {Corna}, \citenamefont {Wuetz}, \citenamefont {Jeon}, \citenamefont {Sheikh}, \citenamefont {Juarez-Hernandez}, \citenamefont {Esparza}, \citenamefont {Rampurawala}, \citenamefont {Carlton}, \citenamefont {Ravikumar}, \citenamefont {Nieva}, \citenamefont {Kim}, \citenamefont {Lee}, \citenamefont {Sammak}, \citenamefont {Scappucci}, \citenamefont {Veldhorst}, \citenamefont {Sebastiano}, \citenamefont {Babaie}, \citenamefont {Pellerano}, \citenamefont {Charbon},\ and\ \citenamefont {Vandersypen}}]{Xue2021}%
  \BibitemOpen
  \bibfield  {author} {\bibinfo {author} {\bibfnamefont {X.}~\bibnamefont {Xue}}, \bibinfo {author} {\bibfnamefont {B.}~\bibnamefont {Patra}}, \bibinfo {author} {\bibfnamefont {J.~P.~G.}\ \bibnamefont {van Dijk}}, \bibinfo {author} {\bibfnamefont {N.}~\bibnamefont {Samkharadze}}, \bibinfo {author} {\bibfnamefont {S.}~\bibnamefont {Subramanian}}, \bibinfo {author} {\bibfnamefont {A.}~\bibnamefont {Corna}}, \bibinfo {author} {\bibfnamefont {B.~P.}\ \bibnamefont {Wuetz}}, \bibinfo {author} {\bibfnamefont {C.}~\bibnamefont {Jeon}}, \bibinfo {author} {\bibfnamefont {F.}~\bibnamefont {Sheikh}}, \bibinfo {author} {\bibfnamefont {E.}~\bibnamefont {Juarez-Hernandez}}, \bibinfo {author} {\bibfnamefont {B.~P.}\ \bibnamefont {Esparza}}, \bibinfo {author} {\bibfnamefont {H.}~\bibnamefont {Rampurawala}}, \bibinfo {author} {\bibfnamefont {B.}~\bibnamefont {Carlton}}, \bibinfo {author} {\bibfnamefont {S.}~\bibnamefont {Ravikumar}}, \bibinfo {author} {\bibfnamefont {C.}~\bibnamefont {Nieva}}, \bibinfo {author} {\bibfnamefont
  {S.}~\bibnamefont {Kim}}, \bibinfo {author} {\bibfnamefont {H.-J.}\ \bibnamefont {Lee}}, \bibinfo {author} {\bibfnamefont {A.}~\bibnamefont {Sammak}}, \bibinfo {author} {\bibfnamefont {G.}~\bibnamefont {Scappucci}}, \bibinfo {author} {\bibfnamefont {M.}~\bibnamefont {Veldhorst}}, \bibinfo {author} {\bibfnamefont {F.}~\bibnamefont {Sebastiano}}, \bibinfo {author} {\bibfnamefont {M.}~\bibnamefont {Babaie}}, \bibinfo {author} {\bibfnamefont {S.}~\bibnamefont {Pellerano}}, \bibinfo {author} {\bibfnamefont {E.}~\bibnamefont {Charbon}},\ and\ \bibinfo {author} {\bibfnamefont {L.~M.~K.}\ \bibnamefont {Vandersypen}},\ }\bibfield  {title} {\bibinfo {title} {{CMOS-based cryogenic control of silicon quantum circuits}},\ }\href {https://doi.org/10.1038/s41586-021-03469-4} {\bibfield  {journal} {\bibinfo  {journal} {Nature}\ }\textbf {\bibinfo {volume} {593}},\ \bibinfo {pages} {205} (\bibinfo {year} {2021})}\BibitemShut {NoStop}%
\bibitem [{\citenamefont {Howe}\ \emph {et~al.}(2022)\citenamefont {Howe}, \citenamefont {Castellanos-Beltran}, \citenamefont {Sirois}, \citenamefont {Olaya}, \citenamefont {Biesecker}, \citenamefont {Dresselhaus}, \citenamefont {Benz},\ and\ \citenamefont {Hopkins}}]{CBeltran2022}%
  \BibitemOpen
  \bibfield  {author} {\bibinfo {author} {\bibfnamefont {L.}~\bibnamefont {Howe}}, \bibinfo {author} {\bibfnamefont {M.~A.}\ \bibnamefont {Castellanos-Beltran}}, \bibinfo {author} {\bibfnamefont {A.~J.}\ \bibnamefont {Sirois}}, \bibinfo {author} {\bibfnamefont {D.}~\bibnamefont {Olaya}}, \bibinfo {author} {\bibfnamefont {J.}~\bibnamefont {Biesecker}}, \bibinfo {author} {\bibfnamefont {P.~D.}\ \bibnamefont {Dresselhaus}}, \bibinfo {author} {\bibfnamefont {S.~P.}\ \bibnamefont {Benz}},\ and\ \bibinfo {author} {\bibfnamefont {P.~F.}\ \bibnamefont {Hopkins}},\ }\bibfield  {title} {\bibinfo {title} {Digital control of a superconducting qubit using a josephson pulse generator at 3 k},\ }\href@noop {} {\bibfield  {journal} {\bibinfo  {journal} {PRX Quantum}\ }\textbf {\bibinfo {volume} {3}},\ \bibinfo {pages} {010350} (\bibinfo {year} {2022})}\BibitemShut {NoStop}%
\bibitem [{\citenamefont {Underwood}\ \emph {et~al.}(2023)\citenamefont {Underwood}, \citenamefont {Glick}, \citenamefont {Inoue}, \citenamefont {Frank}, \citenamefont {Timmerwilke}, \citenamefont {Pritchett}, \citenamefont {Chakraborty}, \citenamefont {Tien}, \citenamefont {Yeck}, \citenamefont {Bulzacchelli}, \citenamefont {Baks}, \citenamefont {Rosno}, \citenamefont {Robertazzi}, \citenamefont {Beck}, \citenamefont {Joshi}, \citenamefont {Wisnieff}, \citenamefont {Ramirez}, \citenamefont {Ruedinger}, \citenamefont {Lekuch}, \citenamefont {Gaucher},\ and\ \citenamefont {Friedman}}]{underwood2023using}%
  \BibitemOpen
  \bibfield  {author} {\bibinfo {author} {\bibfnamefont {D.~L.}\ \bibnamefont {Underwood}}, \bibinfo {author} {\bibfnamefont {J.~A.}\ \bibnamefont {Glick}}, \bibinfo {author} {\bibfnamefont {K.}~\bibnamefont {Inoue}}, \bibinfo {author} {\bibfnamefont {D.~J.}\ \bibnamefont {Frank}}, \bibinfo {author} {\bibfnamefont {J.}~\bibnamefont {Timmerwilke}}, \bibinfo {author} {\bibfnamefont {E.}~\bibnamefont {Pritchett}}, \bibinfo {author} {\bibfnamefont {S.}~\bibnamefont {Chakraborty}}, \bibinfo {author} {\bibfnamefont {K.}~\bibnamefont {Tien}}, \bibinfo {author} {\bibfnamefont {M.}~\bibnamefont {Yeck}}, \bibinfo {author} {\bibfnamefont {J.~F.}\ \bibnamefont {Bulzacchelli}}, \bibinfo {author} {\bibfnamefont {C.}~\bibnamefont {Baks}}, \bibinfo {author} {\bibfnamefont {P.}~\bibnamefont {Rosno}}, \bibinfo {author} {\bibfnamefont {R.}~\bibnamefont {Robertazzi}}, \bibinfo {author} {\bibfnamefont {M.}~\bibnamefont {Beck}}, \bibinfo {author} {\bibfnamefont {R.~V.}\ \bibnamefont {Joshi}}, \bibinfo {author} {\bibfnamefont
  {D.}~\bibnamefont {Wisnieff}}, \bibinfo {author} {\bibfnamefont {D.}~\bibnamefont {Ramirez}}, \bibinfo {author} {\bibfnamefont {J.}~\bibnamefont {Ruedinger}}, \bibinfo {author} {\bibfnamefont {S.}~\bibnamefont {Lekuch}}, \bibinfo {author} {\bibfnamefont {B.~P.}\ \bibnamefont {Gaucher}},\ and\ \bibinfo {author} {\bibfnamefont {D.~J.}\ \bibnamefont {Friedman}},\ }\href@noop {} {\bibinfo {title} {Using cryogenic cmos control electronics to enable a two-qubit cross-resonance gate}} (\bibinfo {year} {2023})\BibitemShut {NoStop}%
\bibitem [{\citenamefont {Bardin}\ \emph {et~al.}(2019)\citenamefont {Bardin}, \citenamefont {Jeffrey}, \citenamefont {Lucero}, \citenamefont {Huang}, \citenamefont {Naaman}, \citenamefont {Barends}, \citenamefont {White}, \citenamefont {Giustina}, \citenamefont {Sank}, \citenamefont {Roushan}, \citenamefont {Arya}, \citenamefont {Chiaro}, \citenamefont {Kelly}, \citenamefont {Chen}, \citenamefont {Burkett}, \citenamefont {Chen}, \citenamefont {Dunsworth}, \citenamefont {Fowler}, \citenamefont {Foxen}, \citenamefont {Gidney}, \citenamefont {Graff}, \citenamefont {Klimov}, \citenamefont {Mutus}, \citenamefont {McEwen}, \citenamefont {Megrant}, \citenamefont {Neeley}, \citenamefont {Neill}, \citenamefont {Quintana}, \citenamefont {Vainsencher}, \citenamefont {Neven},\ and\ \citenamefont {Martinis}}]{Bardin2019}%
  \BibitemOpen
  \bibfield  {author} {\bibinfo {author} {\bibfnamefont {J.~C.}\ \bibnamefont {Bardin}}, \bibinfo {author} {\bibfnamefont {E.}~\bibnamefont {Jeffrey}}, \bibinfo {author} {\bibfnamefont {E.}~\bibnamefont {Lucero}}, \bibinfo {author} {\bibfnamefont {T.}~\bibnamefont {Huang}}, \bibinfo {author} {\bibfnamefont {O.}~\bibnamefont {Naaman}}, \bibinfo {author} {\bibfnamefont {R.}~\bibnamefont {Barends}}, \bibinfo {author} {\bibfnamefont {T.}~\bibnamefont {White}}, \bibinfo {author} {\bibfnamefont {M.}~\bibnamefont {Giustina}}, \bibinfo {author} {\bibfnamefont {D.}~\bibnamefont {Sank}}, \bibinfo {author} {\bibfnamefont {P.}~\bibnamefont {Roushan}}, \bibinfo {author} {\bibfnamefont {K.}~\bibnamefont {Arya}}, \bibinfo {author} {\bibfnamefont {B.}~\bibnamefont {Chiaro}}, \bibinfo {author} {\bibfnamefont {J.}~\bibnamefont {Kelly}}, \bibinfo {author} {\bibfnamefont {J.}~\bibnamefont {Chen}}, \bibinfo {author} {\bibfnamefont {B.}~\bibnamefont {Burkett}}, \bibinfo {author} {\bibfnamefont {Y.}~\bibnamefont {Chen}}, \bibinfo
  {author} {\bibfnamefont {A.}~\bibnamefont {Dunsworth}}, \bibinfo {author} {\bibfnamefont {A.}~\bibnamefont {Fowler}}, \bibinfo {author} {\bibfnamefont {B.}~\bibnamefont {Foxen}}, \bibinfo {author} {\bibfnamefont {C.}~\bibnamefont {Gidney}}, \bibinfo {author} {\bibfnamefont {R.}~\bibnamefont {Graff}}, \bibinfo {author} {\bibfnamefont {P.}~\bibnamefont {Klimov}}, \bibinfo {author} {\bibfnamefont {J.}~\bibnamefont {Mutus}}, \bibinfo {author} {\bibfnamefont {M.}~\bibnamefont {McEwen}}, \bibinfo {author} {\bibfnamefont {A.}~\bibnamefont {Megrant}}, \bibinfo {author} {\bibfnamefont {M.}~\bibnamefont {Neeley}}, \bibinfo {author} {\bibfnamefont {C.}~\bibnamefont {Neill}}, \bibinfo {author} {\bibfnamefont {C.}~\bibnamefont {Quintana}}, \bibinfo {author} {\bibfnamefont {A.}~\bibnamefont {Vainsencher}}, \bibinfo {author} {\bibfnamefont {H.}~\bibnamefont {Neven}},\ and\ \bibinfo {author} {\bibfnamefont {J.}~\bibnamefont {Martinis}},\ }\bibfield  {title} {\bibinfo {title} {29.1 a 28nm bulk-cmos 4-to-8ghz ¡2mw cryogenic
  pulse modulator for scalable quantum computing},\ }in\ \href@noop {} {\emph {\bibinfo {booktitle} {2019 IEEE International Solid- State Circuits Conference - (ISSCC)}}}\ (\bibinfo {year} {2019})\ pp.\ \bibinfo {pages} {456--458}\BibitemShut {NoStop}%
\bibitem [{\citenamefont {Pauka}\ \emph {et~al.}(2021)\citenamefont {Pauka}, \citenamefont {Das}, \citenamefont {Kalra}, \citenamefont {Moini}, \citenamefont {Yang}, \citenamefont {Trainer}, \citenamefont {Bousquet}, \citenamefont {Cantaloube}, \citenamefont {Dick}, \citenamefont {Gardner}, \citenamefont {Manfra},\ and\ \citenamefont {Reilly}}]{Pauka2021}%
  \BibitemOpen
  \bibfield  {author} {\bibinfo {author} {\bibfnamefont {S.~J.}\ \bibnamefont {Pauka}}, \bibinfo {author} {\bibfnamefont {K.}~\bibnamefont {Das}}, \bibinfo {author} {\bibfnamefont {R.}~\bibnamefont {Kalra}}, \bibinfo {author} {\bibfnamefont {A.}~\bibnamefont {Moini}}, \bibinfo {author} {\bibfnamefont {Y.}~\bibnamefont {Yang}}, \bibinfo {author} {\bibfnamefont {M.}~\bibnamefont {Trainer}}, \bibinfo {author} {\bibfnamefont {A.}~\bibnamefont {Bousquet}}, \bibinfo {author} {\bibfnamefont {C.}~\bibnamefont {Cantaloube}}, \bibinfo {author} {\bibfnamefont {N.}~\bibnamefont {Dick}}, \bibinfo {author} {\bibfnamefont {G.~C.}\ \bibnamefont {Gardner}}, \bibinfo {author} {\bibfnamefont {M.~J.}\ \bibnamefont {Manfra}},\ and\ \bibinfo {author} {\bibfnamefont {D.~J.}\ \bibnamefont {Reilly}},\ }\bibfield  {title} {\bibinfo {title} {A cryogenic cmos chip for generating control signals for multiple qubits},\ }\href@noop {} {\bibfield  {journal} {\bibinfo  {journal} {Nature Electronics}\ }\textbf {\bibinfo {volume} {4}},\
  \bibinfo {pages} {64} (\bibinfo {year} {2021})}\BibitemShut {NoStop}%
\bibitem [{\citenamefont {Park}\ \emph {et~al.}(2021)\citenamefont {Park}, \citenamefont {Subramanian}, \citenamefont {Lampert}, \citenamefont {Mladenov}, \citenamefont {Klotchkov}, \citenamefont {Kurian}, \citenamefont {Juarez-Hernandez}, \citenamefont {Perez-Esparza}, \citenamefont {Kale}, \citenamefont {Asma~Beevi}, \citenamefont {Premaratne}, \citenamefont {Watson}, \citenamefont {Suzuki}, \citenamefont {Rahman}, \citenamefont {Timbadiya}, \citenamefont {Soni},\ and\ \citenamefont {Pellerano}}]{Park2021}%
  \BibitemOpen
  \bibfield  {author} {\bibinfo {author} {\bibfnamefont {J.-S.}\ \bibnamefont {Park}}, \bibinfo {author} {\bibfnamefont {S.}~\bibnamefont {Subramanian}}, \bibinfo {author} {\bibfnamefont {L.}~\bibnamefont {Lampert}}, \bibinfo {author} {\bibfnamefont {T.}~\bibnamefont {Mladenov}}, \bibinfo {author} {\bibfnamefont {I.}~\bibnamefont {Klotchkov}}, \bibinfo {author} {\bibfnamefont {D.~J.}\ \bibnamefont {Kurian}}, \bibinfo {author} {\bibfnamefont {E.}~\bibnamefont {Juarez-Hernandez}}, \bibinfo {author} {\bibfnamefont {B.}~\bibnamefont {Perez-Esparza}}, \bibinfo {author} {\bibfnamefont {S.~R.}\ \bibnamefont {Kale}}, \bibinfo {author} {\bibfnamefont {K.~T.}\ \bibnamefont {Asma~Beevi}}, \bibinfo {author} {\bibfnamefont {S.}~\bibnamefont {Premaratne}}, \bibinfo {author} {\bibfnamefont {T.}~\bibnamefont {Watson}}, \bibinfo {author} {\bibfnamefont {S.}~\bibnamefont {Suzuki}}, \bibinfo {author} {\bibfnamefont {M.}~\bibnamefont {Rahman}}, \bibinfo {author} {\bibfnamefont {J.~B.}\ \bibnamefont {Timbadiya}}, \bibinfo
  {author} {\bibfnamefont {S.}~\bibnamefont {Soni}},\ and\ \bibinfo {author} {\bibfnamefont {S.}~\bibnamefont {Pellerano}},\ }\bibfield  {title} {\bibinfo {title} {13.1 a fully integrated cryo-cmos soc for qubit control in quantum computers capable of state manipulation, readout and high-speed gate pulsing of spin qubits in intel 22nm ffl finfet technology},\ }in\ \href {https://doi.org/10.1109/ISSCC42613.2021.9365762} {\emph {\bibinfo {booktitle} {2021 IEEE International Solid- State Circuits Conference (ISSCC)}}},\ Vol.~\bibinfo {volume} {64}\ (\bibinfo {year} {2021})\ pp.\ \bibinfo {pages} {208--210}\BibitemShut {NoStop}%
\bibitem [{\citenamefont {Le~Guevel}\ \emph {et~al.}(2020{\natexlab{a}})\citenamefont {Le~Guevel}, \citenamefont {Billiot}, \citenamefont {Cardoso~Paz}, \citenamefont {Tagliaferri}, \citenamefont {De~Franceschi}, \citenamefont {Maurand}, \citenamefont {Cassé}, \citenamefont {Zurita}, \citenamefont {Sanquer}, \citenamefont {Vinet}, \citenamefont {Jehl}, \citenamefont {Jansen},\ and\ \citenamefont {Pillonnet}}]{Guevel2020}%
  \BibitemOpen
  \bibfield  {author} {\bibinfo {author} {\bibfnamefont {L.}~\bibnamefont {Le~Guevel}}, \bibinfo {author} {\bibfnamefont {G.}~\bibnamefont {Billiot}}, \bibinfo {author} {\bibfnamefont {B.}~\bibnamefont {Cardoso~Paz}}, \bibinfo {author} {\bibfnamefont {M.~L.~V.}\ \bibnamefont {Tagliaferri}}, \bibinfo {author} {\bibfnamefont {S.}~\bibnamefont {De~Franceschi}}, \bibinfo {author} {\bibfnamefont {R.}~\bibnamefont {Maurand}}, \bibinfo {author} {\bibfnamefont {M.}~\bibnamefont {Cassé}}, \bibinfo {author} {\bibfnamefont {M.}~\bibnamefont {Zurita}}, \bibinfo {author} {\bibfnamefont {M.}~\bibnamefont {Sanquer}}, \bibinfo {author} {\bibfnamefont {M.}~\bibnamefont {Vinet}}, \bibinfo {author} {\bibfnamefont {X.}~\bibnamefont {Jehl}}, \bibinfo {author} {\bibfnamefont {A.~G.~M.}\ \bibnamefont {Jansen}},\ and\ \bibinfo {author} {\bibfnamefont {G.}~\bibnamefont {Pillonnet}},\ }\bibfield  {title} {\bibinfo {title} {{Low-power transimpedance amplifier for cryogenic integration with quantum devices}},\ }\href@noop {} {\bibfield
   {journal} {\bibinfo  {journal} {Applied Physics Reviews}\ }\textbf {\bibinfo {volume} {7}} (\bibinfo {year} {2020}{\natexlab{a}})},\ \bibinfo {note} {041407}\BibitemShut {NoStop}%
\bibitem [{\citenamefont {Prabowo}\ \emph {et~al.}(2021)\citenamefont {Prabowo}, \citenamefont {Zheng}, \citenamefont {Mehrpoo}, \citenamefont {Patra}, \citenamefont {Harvey-Collard}, \citenamefont {Dijkema}, \citenamefont {Sammak}, \citenamefont {Scappucci}, \citenamefont {Charbon}, \citenamefont {Sebastiano}, \citenamefont {Vandersypen},\ and\ \citenamefont {Babaie}}]{Prabowo2021}%
  \BibitemOpen
  \bibfield  {author} {\bibinfo {author} {\bibfnamefont {B.}~\bibnamefont {Prabowo}}, \bibinfo {author} {\bibfnamefont {G.}~\bibnamefont {Zheng}}, \bibinfo {author} {\bibfnamefont {M.}~\bibnamefont {Mehrpoo}}, \bibinfo {author} {\bibfnamefont {B.}~\bibnamefont {Patra}}, \bibinfo {author} {\bibfnamefont {P.}~\bibnamefont {Harvey-Collard}}, \bibinfo {author} {\bibfnamefont {J.}~\bibnamefont {Dijkema}}, \bibinfo {author} {\bibfnamefont {A.}~\bibnamefont {Sammak}}, \bibinfo {author} {\bibfnamefont {G.}~\bibnamefont {Scappucci}}, \bibinfo {author} {\bibfnamefont {E.}~\bibnamefont {Charbon}}, \bibinfo {author} {\bibfnamefont {F.}~\bibnamefont {Sebastiano}}, \bibinfo {author} {\bibfnamefont {L.~M.~K.}\ \bibnamefont {Vandersypen}},\ and\ \bibinfo {author} {\bibfnamefont {M.}~\bibnamefont {Babaie}},\ }\bibfield  {title} {\bibinfo {title} {13.3 a 6-to-8ghz 0.17mw/qubit cryo-{CMOS} receiver for multiple spin qubit readout in 40nm {CMOS} technology},\ }in\ \href {https://doi.org/10.1109/isscc42613.2021.9365848} {\emph
  {\bibinfo {booktitle} {2021 {IEEE} International Solid- State Circuits Conference ({ISSCC})}}}\ (\bibinfo  {publisher} {{IEEE}},\ \bibinfo {year} {2021})\BibitemShut {NoStop}%
\bibitem [{\citenamefont {Ruffino}\ \emph {et~al.}(2021)\citenamefont {Ruffino}, \citenamefont {Peng}, \citenamefont {Yang}, \citenamefont {Michniewicz}, \citenamefont {Gonzalez-Zalba},\ and\ \citenamefont {Charbon}}]{Ruffino2021}%
  \BibitemOpen
  \bibfield  {author} {\bibinfo {author} {\bibfnamefont {A.}~\bibnamefont {Ruffino}}, \bibinfo {author} {\bibfnamefont {Y.}~\bibnamefont {Peng}}, \bibinfo {author} {\bibfnamefont {T.-Y.}\ \bibnamefont {Yang}}, \bibinfo {author} {\bibfnamefont {J.}~\bibnamefont {Michniewicz}}, \bibinfo {author} {\bibfnamefont {M.~F.}\ \bibnamefont {Gonzalez-Zalba}},\ and\ \bibinfo {author} {\bibfnamefont {E.}~\bibnamefont {Charbon}},\ }\bibfield  {title} {\bibinfo {title} {13.2 a fully-integrated 40-nm 5-6.5 ghz cryo-cmos system-on-chip with i/q receiver and frequency synthesizer for scalable multiplexed readout of quantum dots},\ }in\ \href {https://doi.org/10.1109/ISSCC42613.2021.9365758} {\emph {\bibinfo {booktitle} {2021 IEEE International Solid- State Circuits Conference (ISSCC)}}},\ Vol.~\bibinfo {volume} {64}\ (\bibinfo {year} {2021})\ pp.\ \bibinfo {pages} {210--212}\BibitemShut {NoStop}%
\bibitem [{\citenamefont {Guevel}\ \emph {et~al.}(2020)\citenamefont {Guevel}, \citenamefont {Billiot}, \citenamefont {Jehl}, \citenamefont {De~Franceschi}, \citenamefont {Zurita}, \citenamefont {Thonnart}, \citenamefont {Vinet}, \citenamefont {Sanquer}, \citenamefont {Maurand}, \citenamefont {Jansen},\ and\ \citenamefont {Pillonnet}}]{Guevel2020_2}%
  \BibitemOpen
  \bibfield  {author} {\bibinfo {author} {\bibfnamefont {L.~L.}\ \bibnamefont {Guevel}}, \bibinfo {author} {\bibfnamefont {G.}~\bibnamefont {Billiot}}, \bibinfo {author} {\bibfnamefont {X.}~\bibnamefont {Jehl}}, \bibinfo {author} {\bibfnamefont {S.}~\bibnamefont {De~Franceschi}}, \bibinfo {author} {\bibfnamefont {M.}~\bibnamefont {Zurita}}, \bibinfo {author} {\bibfnamefont {Y.}~\bibnamefont {Thonnart}}, \bibinfo {author} {\bibfnamefont {M.}~\bibnamefont {Vinet}}, \bibinfo {author} {\bibfnamefont {M.}~\bibnamefont {Sanquer}}, \bibinfo {author} {\bibfnamefont {R.}~\bibnamefont {Maurand}}, \bibinfo {author} {\bibfnamefont {A.~G.~M.}\ \bibnamefont {Jansen}},\ and\ \bibinfo {author} {\bibfnamefont {G.}~\bibnamefont {Pillonnet}},\ }\bibfield  {title} {\bibinfo {title} {19.2 a 110mk 295µw 28nm fdsoi cmos quantum integrated circuit with a 2.8ghz excitation and na current sensing of an on-chip double quantum dot},\ }in\ \href@noop {} {\emph {\bibinfo {booktitle} {2020 IEEE International Solid- State Circuits
  Conference - (ISSCC)}}}\ (\bibinfo {year} {2020})\ pp.\ \bibinfo {pages} {306--308}\BibitemShut {NoStop}%
\bibitem [{\citenamefont {Triantopoulos}\ \emph {et~al.}(2019)\citenamefont {Triantopoulos}, \citenamefont {Cassé}, \citenamefont {Barraud}, \citenamefont {Haendler}, \citenamefont {Vincent}, \citenamefont {Vinet}, \citenamefont {Gaillard},\ and\ \citenamefont {Ghibaudo}}]{Triantopoulos2019}%
  \BibitemOpen
  \bibfield  {author} {\bibinfo {author} {\bibfnamefont {K.}~\bibnamefont {Triantopoulos}}, \bibinfo {author} {\bibfnamefont {M.}~\bibnamefont {Cassé}}, \bibinfo {author} {\bibfnamefont {S.}~\bibnamefont {Barraud}}, \bibinfo {author} {\bibfnamefont {S.}~\bibnamefont {Haendler}}, \bibinfo {author} {\bibfnamefont {E.}~\bibnamefont {Vincent}}, \bibinfo {author} {\bibfnamefont {M.}~\bibnamefont {Vinet}}, \bibinfo {author} {\bibfnamefont {F.}~\bibnamefont {Gaillard}},\ and\ \bibinfo {author} {\bibfnamefont {G.}~\bibnamefont {Ghibaudo}},\ }\bibfield  {title} {\bibinfo {title} {Self-heating effect in fdsoi transistors down to cryogenic operation at 4.2 k},\ }\href {https://doi.org/10.1109/TED.2019.2919924} {\bibfield  {journal} {\bibinfo  {journal} {IEEE Transactions on Electron Devices}\ }\textbf {\bibinfo {volume} {66}},\ \bibinfo {pages} {3498} (\bibinfo {year} {2019})}\BibitemShut {NoStop}%
\bibitem [{\citenamefont {T~Hart}\ \emph {et~al.}(2021)\citenamefont {T~Hart}, \citenamefont {Babaie}, \citenamefont {Vladimirescu},\ and\ \citenamefont {Sebastiano}}]{Hart2021}%
  \BibitemOpen
  \bibfield  {author} {\bibinfo {author} {\bibfnamefont {P.~A.}\ \bibnamefont {T~Hart}}, \bibinfo {author} {\bibfnamefont {M.}~\bibnamefont {Babaie}}, \bibinfo {author} {\bibfnamefont {A.}~\bibnamefont {Vladimirescu}},\ and\ \bibinfo {author} {\bibfnamefont {F.}~\bibnamefont {Sebastiano}},\ }\bibfield  {title} {\bibinfo {title} {Characterization and modeling of self-heating in nanometer bulk-cmos at cryogenic temperatures},\ }\href {https://doi.org/10.1109/JEDS.2021.3116975} {\bibfield  {journal} {\bibinfo  {journal} {IEEE Journal of the Electron Devices Society}\ }\textbf {\bibinfo {volume} {9}},\ \bibinfo {pages} {891} (\bibinfo {year} {2021})}\BibitemShut {NoStop}%
\bibitem [{\citenamefont {Petit}\ \emph {et~al.}(2020)\citenamefont {Petit}, \citenamefont {Eenink}, \citenamefont {Russ}, \citenamefont {Lawrie}, \citenamefont {Hendrickx}, \citenamefont {Philips}, \citenamefont {Clarke}, \citenamefont {Vandersypen},\ and\ \citenamefont {Veldhorst}}]{Petit2020}%
  \BibitemOpen
  \bibfield  {author} {\bibinfo {author} {\bibfnamefont {L.}~\bibnamefont {Petit}}, \bibinfo {author} {\bibfnamefont {H.~G.~J.}\ \bibnamefont {Eenink}}, \bibinfo {author} {\bibfnamefont {M.}~\bibnamefont {Russ}}, \bibinfo {author} {\bibfnamefont {W.~I.~L.}\ \bibnamefont {Lawrie}}, \bibinfo {author} {\bibfnamefont {N.~W.}\ \bibnamefont {Hendrickx}}, \bibinfo {author} {\bibfnamefont {S.~G.~J.}\ \bibnamefont {Philips}}, \bibinfo {author} {\bibfnamefont {J.~S.}\ \bibnamefont {Clarke}}, \bibinfo {author} {\bibfnamefont {L.~M.~K.}\ \bibnamefont {Vandersypen}},\ and\ \bibinfo {author} {\bibfnamefont {M.}~\bibnamefont {Veldhorst}},\ }\bibfield  {title} {\bibinfo {title} {Universal quantum logic in hot silicon qubits},\ }\href@noop {} {\bibfield  {journal} {\bibinfo  {journal} {Nature}\ }\textbf {\bibinfo {volume} {580}},\ \bibinfo {pages} {355} (\bibinfo {year} {2020})}\BibitemShut {NoStop}%
\bibitem [{\citenamefont {Yang}\ \emph {et~al.}(2020)\citenamefont {Yang}, \citenamefont {Leon}, \citenamefont {Hwang}, \citenamefont {Saraiva}, \citenamefont {Tanttu}, \citenamefont {Huang}, \citenamefont {Camirand~Lemyre}, \citenamefont {Chan}, \citenamefont {Tan}, \citenamefont {Hudson}, \citenamefont {Itoh}, \citenamefont {Morello}, \citenamefont {Pioro-Ladri{\`e}re}, \citenamefont {Laucht},\ and\ \citenamefont {Dzurak}}]{Yang2020}%
  \BibitemOpen
  \bibfield  {author} {\bibinfo {author} {\bibfnamefont {C.~H.}\ \bibnamefont {Yang}}, \bibinfo {author} {\bibfnamefont {R.~C.~C.}\ \bibnamefont {Leon}}, \bibinfo {author} {\bibfnamefont {J.~C.~C.}\ \bibnamefont {Hwang}}, \bibinfo {author} {\bibfnamefont {A.}~\bibnamefont {Saraiva}}, \bibinfo {author} {\bibfnamefont {T.}~\bibnamefont {Tanttu}}, \bibinfo {author} {\bibfnamefont {W.}~\bibnamefont {Huang}}, \bibinfo {author} {\bibfnamefont {J.}~\bibnamefont {Camirand~Lemyre}}, \bibinfo {author} {\bibfnamefont {K.~W.}\ \bibnamefont {Chan}}, \bibinfo {author} {\bibfnamefont {K.~Y.}\ \bibnamefont {Tan}}, \bibinfo {author} {\bibfnamefont {F.~E.}\ \bibnamefont {Hudson}}, \bibinfo {author} {\bibfnamefont {K.~M.}\ \bibnamefont {Itoh}}, \bibinfo {author} {\bibfnamefont {A.}~\bibnamefont {Morello}}, \bibinfo {author} {\bibfnamefont {M.}~\bibnamefont {Pioro-Ladri{\`e}re}}, \bibinfo {author} {\bibfnamefont {A.}~\bibnamefont {Laucht}},\ and\ \bibinfo {author} {\bibfnamefont {A.~S.}\ \bibnamefont {Dzurak}},\ }\bibfield
  {title} {\bibinfo {title} {Operation of a silicon quantum processor unit cell above one kelvin},\ }\href@noop {} {\bibfield  {journal} {\bibinfo  {journal} {Nature}\ }\textbf {\bibinfo {volume} {580}},\ \bibinfo {pages} {350} (\bibinfo {year} {2020})}\BibitemShut {NoStop}%
\bibitem [{\citenamefont {Undseth}\ \emph {et~al.}(2023)\citenamefont {Undseth}, \citenamefont {Pietx-Casas}, \citenamefont {Raymenants}, \citenamefont {Mehmandoost}, \citenamefont {Mądzik}, \citenamefont {Philips}, \citenamefont {de~Snoo}, \citenamefont {Michalak}, \citenamefont {Amitonov}, \citenamefont {Tryputen}, \citenamefont {Wuetz}, \citenamefont {Fezzi}, \citenamefont {Esposti}, \citenamefont {Sammak}, \citenamefont {Scappucci},\ and\ \citenamefont {Vandersypen}}]{undseth2023}%
  \BibitemOpen
  \bibfield  {author} {\bibinfo {author} {\bibfnamefont {B.}~\bibnamefont {Undseth}}, \bibinfo {author} {\bibfnamefont {O.}~\bibnamefont {Pietx-Casas}}, \bibinfo {author} {\bibfnamefont {E.}~\bibnamefont {Raymenants}}, \bibinfo {author} {\bibfnamefont {M.}~\bibnamefont {Mehmandoost}}, \bibinfo {author} {\bibfnamefont {M.~T.}\ \bibnamefont {Mądzik}}, \bibinfo {author} {\bibfnamefont {S.~G.~J.}\ \bibnamefont {Philips}}, \bibinfo {author} {\bibfnamefont {S.~L.}\ \bibnamefont {de~Snoo}}, \bibinfo {author} {\bibfnamefont {D.~J.}\ \bibnamefont {Michalak}}, \bibinfo {author} {\bibfnamefont {S.~V.}\ \bibnamefont {Amitonov}}, \bibinfo {author} {\bibfnamefont {L.}~\bibnamefont {Tryputen}}, \bibinfo {author} {\bibfnamefont {B.~P.}\ \bibnamefont {Wuetz}}, \bibinfo {author} {\bibfnamefont {V.}~\bibnamefont {Fezzi}}, \bibinfo {author} {\bibfnamefont {D.~D.}\ \bibnamefont {Esposti}}, \bibinfo {author} {\bibfnamefont {A.}~\bibnamefont {Sammak}}, \bibinfo {author} {\bibfnamefont {G.}~\bibnamefont {Scappucci}},\ and\ \bibinfo
  {author} {\bibfnamefont {L.~M.~K.}\ \bibnamefont {Vandersypen}},\ }\href@noop {} {\bibinfo {title} {Hotter is easier: unexpected temperature dependence of spin qubit frequencies}} (\bibinfo {year} {2023}),\ \Eprint {https://arxiv.org/abs/2304.12984} {arXiv:2304.12984} \BibitemShut {NoStop}%
\bibitem [{\citenamefont {Noah}\ \emph {et~al.}(2023)\citenamefont {Noah}, \citenamefont {Swift}, \citenamefont {de~Kruijf}, \citenamefont {Gomez-Saiz}, \citenamefont {Morton},\ and\ \citenamefont {Gonzalez-Zalba}}]{noah2023cmos}%
  \BibitemOpen
  \bibfield  {author} {\bibinfo {author} {\bibfnamefont {G.~M.}\ \bibnamefont {Noah}}, \bibinfo {author} {\bibfnamefont {T.}~\bibnamefont {Swift}}, \bibinfo {author} {\bibfnamefont {M.}~\bibnamefont {de~Kruijf}}, \bibinfo {author} {\bibfnamefont {A.}~\bibnamefont {Gomez-Saiz}}, \bibinfo {author} {\bibfnamefont {J.~J.~L.}\ \bibnamefont {Morton}},\ and\ \bibinfo {author} {\bibfnamefont {M.~F.}\ \bibnamefont {Gonzalez-Zalba}},\ }\href@noop {} {\bibinfo {title} {Cmos on-chip thermometry at deep cryogenic temperatures}} (\bibinfo {year} {2023}),\ \Eprint {https://arxiv.org/abs/2308.00392} {arXiv:2308.00392 [physics.app-ph]} \BibitemShut {NoStop}%
\bibitem [{\citenamefont {Kouwenhoven}\ and\ \citenamefont {et~al.}(2001)}]{Kouwenhoven2001}%
  \BibitemOpen
  \bibfield  {author} {\bibinfo {author} {\bibfnamefont {L.~P.}\ \bibnamefont {Kouwenhoven}}\ and\ \bibinfo {author} {\bibnamefont {et~al.}},\ }\bibfield  {title} {\bibinfo {title} {Few-electron quantum dots},\ }\href {http://stacks.iop.org/0034-4885/64/i=6/a=201} {\bibfield  {journal} {\bibinfo  {journal} {Reports on Progress in Physics}\ }\textbf {\bibinfo {volume} {64}},\ \bibinfo {pages} {701} (\bibinfo {year} {2001})}\BibitemShut {NoStop}%
\bibitem [{\citenamefont {Gonzalez-Zalba}\ \emph {et~al.}(2015)\citenamefont {Gonzalez-Zalba}, \citenamefont {Barraud}, \citenamefont {Ferguson},\ and\ \citenamefont {Betz}}]{Gonzalez-Zalba2015}%
  \BibitemOpen
  \bibfield  {author} {\bibinfo {author} {\bibfnamefont {M.~F.}\ \bibnamefont {Gonzalez-Zalba}}, \bibinfo {author} {\bibfnamefont {S.}~\bibnamefont {Barraud}}, \bibinfo {author} {\bibfnamefont {A.}~\bibnamefont {Ferguson}},\ and\ \bibinfo {author} {\bibfnamefont {A.~C.}\ \bibnamefont {Betz}},\ }\bibfield  {title} {\bibinfo {title} {Probing the limits of gate-based charge sensing},\ }\href@noop {} {\bibfield  {journal} {\bibinfo  {journal} {Nat Commun}\ }\textbf {\bibinfo {volume} {6}},\ \bibinfo {pages} {6084} (\bibinfo {year} {2015})}\BibitemShut {NoStop}%
\bibitem [{\citenamefont {Houten}\ \emph {et~al.}(1992)\citenamefont {Houten}, \citenamefont {Beenakker},\ and\ \citenamefont {Staring}}]{Houten1992}%
  \BibitemOpen
  \bibfield  {author} {\bibinfo {author} {\bibfnamefont {H.~V.}\ \bibnamefont {Houten}}, \bibinfo {author} {\bibfnamefont {C.~W.~J.}\ \bibnamefont {Beenakker}},\ and\ \bibinfo {author} {\bibfnamefont {A.~A.~M.}\ \bibnamefont {Staring}},\ }\bibfield  {title} {\bibinfo {title} {Coulomb-blockade oscillations in semiconductor nanostructures},\ }in\ \href {https://doi.org/10.1007/978-1-4757-2166-9_5} {\emph {\bibinfo {booktitle} {{NATO} {ASI} Series}}}\ (\bibinfo  {publisher} {Springer {US}},\ \bibinfo {year} {1992})\ pp.\ \bibinfo {pages} {167--216}\BibitemShut {NoStop}%
\bibitem [{\citenamefont {Maradan}\ \emph {et~al.}(2014)\citenamefont {Maradan}, \citenamefont {Casparis}, \citenamefont {Liu}, \citenamefont {Biesinger}, \citenamefont {Scheller}, \citenamefont {Zumbühl}, \citenamefont {Zimmerman},\ and\ \citenamefont {Gossard}}]{Maradan2014}%
  \BibitemOpen
  \bibfield  {author} {\bibinfo {author} {\bibfnamefont {D.}~\bibnamefont {Maradan}}, \bibinfo {author} {\bibfnamefont {L.}~\bibnamefont {Casparis}}, \bibinfo {author} {\bibfnamefont {T.-M.}\ \bibnamefont {Liu}}, \bibinfo {author} {\bibfnamefont {D.~E.~F.}\ \bibnamefont {Biesinger}}, \bibinfo {author} {\bibfnamefont {C.~P.}\ \bibnamefont {Scheller}}, \bibinfo {author} {\bibfnamefont {D.~M.}\ \bibnamefont {Zumbühl}}, \bibinfo {author} {\bibfnamefont {J.~D.}\ \bibnamefont {Zimmerman}},\ and\ \bibinfo {author} {\bibfnamefont {A.~C.}\ \bibnamefont {Gossard}},\ }\bibfield  {title} {\bibinfo {title} {{GaAs} quantum dot thermometry using direct transport and charge sensing},\ }\href {https://doi.org/10.1007/s10909-014-1169-6} {\bibfield  {journal} {\bibinfo  {journal} {Journal of Low Temperature Physics}\ }\textbf {\bibinfo {volume} {175}},\ \bibinfo {pages} {784} (\bibinfo {year} {2014})}\BibitemShut {NoStop}%
\bibitem [{\citenamefont {Escott}\ \emph {et~al.}(2010)\citenamefont {Escott}, \citenamefont {Zwanenburg},\ and\ \citenamefont {Morello}}]{Escott_2010}%
  \BibitemOpen
  \bibfield  {author} {\bibinfo {author} {\bibfnamefont {C.~C.}\ \bibnamefont {Escott}}, \bibinfo {author} {\bibfnamefont {F.~A.}\ \bibnamefont {Zwanenburg}},\ and\ \bibinfo {author} {\bibfnamefont {A.}~\bibnamefont {Morello}},\ }\bibfield  {title} {\bibinfo {title} {Resonant tunnelling features in quantum dots},\ }\href {https://doi.org/10.1088/0957-4484/21/27/274018} {\bibfield  {journal} {\bibinfo  {journal} {Nanotechnology}\ }\textbf {\bibinfo {volume} {21}},\ \bibinfo {pages} {274018} (\bibinfo {year} {2010})}\BibitemShut {NoStop}%
\bibitem [{\citenamefont {Iftikhar}\ \emph {et~al.}(2016)\citenamefont {Iftikhar}, \citenamefont {Anthore}, \citenamefont {Jezouin}, \citenamefont {Parmentier}, \citenamefont {Jin}, \citenamefont {Cavanna}, \citenamefont {Ouerghi}, \citenamefont {Gennser},\ and\ \citenamefont {Pierre}}]{Iftikhar2016}%
  \BibitemOpen
  \bibfield  {author} {\bibinfo {author} {\bibfnamefont {Z.}~\bibnamefont {Iftikhar}}, \bibinfo {author} {\bibfnamefont {A.}~\bibnamefont {Anthore}}, \bibinfo {author} {\bibfnamefont {S.}~\bibnamefont {Jezouin}}, \bibinfo {author} {\bibfnamefont {F.~D.}\ \bibnamefont {Parmentier}}, \bibinfo {author} {\bibfnamefont {Y.}~\bibnamefont {Jin}}, \bibinfo {author} {\bibfnamefont {A.}~\bibnamefont {Cavanna}}, \bibinfo {author} {\bibfnamefont {A.}~\bibnamefont {Ouerghi}}, \bibinfo {author} {\bibfnamefont {U.}~\bibnamefont {Gennser}},\ and\ \bibinfo {author} {\bibfnamefont {F.}~\bibnamefont {Pierre}},\ }\bibfield  {title} {\bibinfo {title} {Primary thermometry triad at 6 mk in mesoscopic circuits},\ }\href {http://dx.doi.org/10.1038/ncomms12908} {\bibfield  {journal} {\bibinfo  {journal} {Nature Communications}\ }\textbf {\bibinfo {volume} {7}},\ \bibinfo {pages} {12908 EP } (\bibinfo {year} {2016})},\ \bibinfo {note} {article}\BibitemShut {NoStop}%
\bibitem [{\citenamefont {Ahmed}\ \emph {et~al.}(2018)\citenamefont {Ahmed}, \citenamefont {Chatterjee}, \citenamefont {Barraud}, \citenamefont {Morton}, \citenamefont {Haigh},\ and\ \citenamefont {Gonzalez-Zalba}}]{Ahmed2018b}%
  \BibitemOpen
  \bibfield  {author} {\bibinfo {author} {\bibfnamefont {I.}~\bibnamefont {Ahmed}}, \bibinfo {author} {\bibfnamefont {A.}~\bibnamefont {Chatterjee}}, \bibinfo {author} {\bibfnamefont {S.}~\bibnamefont {Barraud}}, \bibinfo {author} {\bibfnamefont {J.~J.~L.}\ \bibnamefont {Morton}}, \bibinfo {author} {\bibfnamefont {J.~A.}\ \bibnamefont {Haigh}},\ and\ \bibinfo {author} {\bibfnamefont {M.~F.}\ \bibnamefont {Gonzalez-Zalba}},\ }\bibfield  {title} {\bibinfo {title} {Primary thermometry of a single reservoir using cyclic electron tunneling to a quantum dot},\ }\href {https://doi.org/10.1038/s42005-018-0066-8} {\bibfield  {journal} {\bibinfo  {journal} {Communications Physics}\ }\textbf {\bibinfo {volume} {1}},\ \bibinfo {pages} {66} (\bibinfo {year} {2018})}\BibitemShut {NoStop}%
\bibitem [{\citenamefont {Kranz}\ \emph {et~al.}(2020)\citenamefont {Kranz}, \citenamefont {Gorman}, \citenamefont {Thorgrimsson}, \citenamefont {He}, \citenamefont {Keith}, \citenamefont {Keizer},\ and\ \citenamefont {Simmons}}]{Kranz2020}%
  \BibitemOpen
  \bibfield  {author} {\bibinfo {author} {\bibfnamefont {L.}~\bibnamefont {Kranz}}, \bibinfo {author} {\bibfnamefont {S.~K.}\ \bibnamefont {Gorman}}, \bibinfo {author} {\bibfnamefont {B.}~\bibnamefont {Thorgrimsson}}, \bibinfo {author} {\bibfnamefont {Y.}~\bibnamefont {He}}, \bibinfo {author} {\bibfnamefont {D.}~\bibnamefont {Keith}}, \bibinfo {author} {\bibfnamefont {J.~G.}\ \bibnamefont {Keizer}},\ and\ \bibinfo {author} {\bibfnamefont {M.~Y.}\ \bibnamefont {Simmons}},\ }\bibfield  {title} {\bibinfo {title} {Exploiting a single-crystal environment to minimize the charge noise on qubits in silicon},\ }\href@noop {} {\bibfield  {journal} {\bibinfo  {journal} {Advanced Materials}\ }\textbf {\bibinfo {volume} {32}},\ \bibinfo {pages} {2003361} (\bibinfo {year} {2020})}\BibitemShut {NoStop}%
\bibitem [{\citenamefont {Spence}\ \emph {et~al.}(2022)\citenamefont {Spence}, \citenamefont {Cardoso-Paz}, \citenamefont {Michal}, \citenamefont {Chanrion}, \citenamefont {Niegemann}, \citenamefont {Jadot}, \citenamefont {Mortemousque}, \citenamefont {Klemt}, \citenamefont {Thiney}, \citenamefont {Bertrand}, \citenamefont {Hutin}, \citenamefont {Bäuerle}, \citenamefont {Balestro}, \citenamefont {Vinet}, \citenamefont {Niquet}, \citenamefont {Meunier},\ and\ \citenamefont {Urdampilleta}}]{Spence2022}%
  \BibitemOpen
  \bibfield  {author} {\bibinfo {author} {\bibfnamefont {C.}~\bibnamefont {Spence}}, \bibinfo {author} {\bibfnamefont {B.}~\bibnamefont {Cardoso-Paz}}, \bibinfo {author} {\bibfnamefont {V.}~\bibnamefont {Michal}}, \bibinfo {author} {\bibfnamefont {E.}~\bibnamefont {Chanrion}}, \bibinfo {author} {\bibfnamefont {D.~J.}\ \bibnamefont {Niegemann}}, \bibinfo {author} {\bibfnamefont {B.}~\bibnamefont {Jadot}}, \bibinfo {author} {\bibfnamefont {P.-A.}\ \bibnamefont {Mortemousque}}, \bibinfo {author} {\bibfnamefont {B.}~\bibnamefont {Klemt}}, \bibinfo {author} {\bibfnamefont {V.}~\bibnamefont {Thiney}}, \bibinfo {author} {\bibfnamefont {B.}~\bibnamefont {Bertrand}}, \bibinfo {author} {\bibfnamefont {L.}~\bibnamefont {Hutin}}, \bibinfo {author} {\bibfnamefont {C.}~\bibnamefont {Bäuerle}}, \bibinfo {author} {\bibfnamefont {F.}~\bibnamefont {Balestro}}, \bibinfo {author} {\bibfnamefont {M.}~\bibnamefont {Vinet}}, \bibinfo {author} {\bibfnamefont {Y.-M.}\ \bibnamefont {Niquet}}, \bibinfo {author} {\bibfnamefont
  {T.}~\bibnamefont {Meunier}},\ and\ \bibinfo {author} {\bibfnamefont {M.}~\bibnamefont {Urdampilleta}},\ }\href@noop {} {\bibinfo {title} {Probing charge noise in few electron cmos quantum dots}} (\bibinfo {year} {2022}),\ \Eprint {https://arxiv.org/abs/2209.01853} {arXiv:2209.01853 [cond-mat.mes-hall]} \BibitemShut {NoStop}%
\bibitem [{\citenamefont {Richardson}(1988)}]{Richardson1988}%
  \BibitemOpen
  \bibfield  {author} {\bibinfo {author} {\bibfnamefont {R.}~\bibnamefont {Richardson}},\ }\href@noop {} {\emph {\bibinfo {title} {Experimental Techniques In Condensed Matter Physics At Low Temperatures}}}\ (\bibinfo  {publisher} {CRC Press},\ \bibinfo {year} {1988})\BibitemShut {NoStop}%
\bibitem [{\citenamefont {Duthil}(2014)}]{Duthil2014}%
  \BibitemOpen
  \bibfield  {author} {\bibinfo {author} {\bibfnamefont {P.}~\bibnamefont {Duthil}},\ }\bibfield  {title} {\bibinfo {title} {{Material Properties at Low Temperature}},\ }\href {https://doi.org/10.5170/CERN-2014-005.77} {\ ,\ \bibinfo {pages} {77} (\bibinfo {year} {2014})},\ \bibinfo {note} {comments: 18 pages, contribution to the CAS-CERN Accelerator School: Superconductivity for Accelerators, Erice, Italy, 24 April - 4 May 2013, edited by R. Bailey},\ \Eprint {https://arxiv.org/abs/1501.07100} {arXiv:1501.07100} \BibitemShut {NoStop}%
\bibitem [{\citenamefont {Acharya}\ \emph {et~al.}(2023)\citenamefont {Acharya}, \citenamefont {Brebels}, \citenamefont {Grill}, \citenamefont {Verjauw}, \citenamefont {Ivanov}, \citenamefont {Lozano}, \citenamefont {Wan}, \citenamefont {Van~Damme}, \citenamefont {Vadiraj}, \citenamefont {Mongillo}, \citenamefont {Govoreanu}, \citenamefont {Craninckx}, \citenamefont {Radu}, \citenamefont {De~Greve}, \citenamefont {Gielen}, \citenamefont {Catthoor},\ and\ \citenamefont {Poto{\v{c}}nik}}]{Acharya2023}%
  \BibitemOpen
  \bibfield  {author} {\bibinfo {author} {\bibfnamefont {R.}~\bibnamefont {Acharya}}, \bibinfo {author} {\bibfnamefont {S.}~\bibnamefont {Brebels}}, \bibinfo {author} {\bibfnamefont {A.}~\bibnamefont {Grill}}, \bibinfo {author} {\bibfnamefont {J.}~\bibnamefont {Verjauw}}, \bibinfo {author} {\bibfnamefont {T.}~\bibnamefont {Ivanov}}, \bibinfo {author} {\bibfnamefont {D.~P.}\ \bibnamefont {Lozano}}, \bibinfo {author} {\bibfnamefont {D.}~\bibnamefont {Wan}}, \bibinfo {author} {\bibfnamefont {J.}~\bibnamefont {Van~Damme}}, \bibinfo {author} {\bibfnamefont {A.~M.}\ \bibnamefont {Vadiraj}}, \bibinfo {author} {\bibfnamefont {M.}~\bibnamefont {Mongillo}}, \bibinfo {author} {\bibfnamefont {B.}~\bibnamefont {Govoreanu}}, \bibinfo {author} {\bibfnamefont {J.}~\bibnamefont {Craninckx}}, \bibinfo {author} {\bibfnamefont {I.~P.}\ \bibnamefont {Radu}}, \bibinfo {author} {\bibfnamefont {K.}~\bibnamefont {De~Greve}}, \bibinfo {author} {\bibfnamefont {G.}~\bibnamefont {Gielen}}, \bibinfo {author} {\bibfnamefont
  {F.}~\bibnamefont {Catthoor}},\ and\ \bibinfo {author} {\bibfnamefont {A.}~\bibnamefont {Poto{\v{c}}nik}},\ }\bibfield  {title} {\bibinfo {title} {Multiplexed superconducting qubit control at millikelvin temperatures with a low-power cryo-cmos multiplexer},\ }\href@noop {} {\bibfield  {journal} {\bibinfo  {journal} {Nat Electron}\ } (\bibinfo {year} {2023})}\BibitemShut {NoStop}%
\bibitem [{\citenamefont {Le~Guevel}\ \emph {et~al.}(2020{\natexlab{b}})\citenamefont {Le~Guevel}, \citenamefont {Billiot}, \citenamefont {Cardoso~Paz}, \citenamefont {Tagliaferri}, \citenamefont {De~Franceschi}, \citenamefont {Maurand}, \citenamefont {Cassé}, \citenamefont {Zurita}, \citenamefont {Sanquer}, \citenamefont {Vinet}, \citenamefont {Jehl}, \citenamefont {Jansen},\ and\ \citenamefont {Pillonnet}}]{LeGuevel2020}%
  \BibitemOpen
  \bibfield  {author} {\bibinfo {author} {\bibfnamefont {L.}~\bibnamefont {Le~Guevel}}, \bibinfo {author} {\bibfnamefont {G.}~\bibnamefont {Billiot}}, \bibinfo {author} {\bibfnamefont {B.}~\bibnamefont {Cardoso~Paz}}, \bibinfo {author} {\bibfnamefont {M.~L.~V.}\ \bibnamefont {Tagliaferri}}, \bibinfo {author} {\bibfnamefont {S.}~\bibnamefont {De~Franceschi}}, \bibinfo {author} {\bibfnamefont {R.}~\bibnamefont {Maurand}}, \bibinfo {author} {\bibfnamefont {M.}~\bibnamefont {Cassé}}, \bibinfo {author} {\bibfnamefont {M.}~\bibnamefont {Zurita}}, \bibinfo {author} {\bibfnamefont {M.}~\bibnamefont {Sanquer}}, \bibinfo {author} {\bibfnamefont {M.}~\bibnamefont {Vinet}}, \bibinfo {author} {\bibfnamefont {X.}~\bibnamefont {Jehl}}, \bibinfo {author} {\bibfnamefont {A.~G.~M.}\ \bibnamefont {Jansen}},\ and\ \bibinfo {author} {\bibfnamefont {G.}~\bibnamefont {Pillonnet}},\ }\bibfield  {title} {\bibinfo {title} {{Low-power transimpedance amplifier for cryogenic integration with quantum devices}},\ }\href@noop {} {\bibfield
   {journal} {\bibinfo  {journal} {Applied Physics Reviews}\ }\textbf {\bibinfo {volume} {7}},\ \bibinfo {pages} {041407} (\bibinfo {year} {2020}{\natexlab{b}})}\BibitemShut {NoStop}%
\bibitem [{\citenamefont {{Le Guevel}}\ \emph {et~al.}(2023)\citenamefont {{Le Guevel}}, \citenamefont {Billiot}, \citenamefont {{De Franceschi}}, \citenamefont {Morel}, \citenamefont {Jehl}, \citenamefont {Jansen},\ and\ \citenamefont {Pillonnet}}]{LEGUEVEL2023}%
  \BibitemOpen
  \bibfield  {author} {\bibinfo {author} {\bibfnamefont {L.}~\bibnamefont {{Le Guevel}}}, \bibinfo {author} {\bibfnamefont {G.}~\bibnamefont {Billiot}}, \bibinfo {author} {\bibfnamefont {S.}~\bibnamefont {{De Franceschi}}}, \bibinfo {author} {\bibfnamefont {A.}~\bibnamefont {Morel}}, \bibinfo {author} {\bibfnamefont {X.}~\bibnamefont {Jehl}}, \bibinfo {author} {\bibfnamefont {A.}~\bibnamefont {Jansen}},\ and\ \bibinfo {author} {\bibfnamefont {G.}~\bibnamefont {Pillonnet}},\ }\bibfield  {title} {\bibinfo {title} {Impedancemetry of multiplexed quantum devices using an on-chip cryogenic cmos active inductor},\ }\href {https://doi.org/https://doi.org/10.1016/j.chip.2023.100068} {\bibfield  {journal} {\bibinfo  {journal} {Chip}\ ,\ \bibinfo {pages} {100068}} (\bibinfo {year} {2023})}\BibitemShut {NoStop}%
\bibitem [{\citenamefont {Vigneau}\ \emph {et~al.}(2023)\citenamefont {Vigneau}, \citenamefont {Fedele}, \citenamefont {Chatterjee}, \citenamefont {Reilly}, \citenamefont {Kuemmeth}, \citenamefont {Gonzalez-Zalba}, \citenamefont {Laird},\ and\ \citenamefont {Ares}}]{vigneau2022}%
  \BibitemOpen
  \bibfield  {author} {\bibinfo {author} {\bibfnamefont {F.}~\bibnamefont {Vigneau}}, \bibinfo {author} {\bibfnamefont {F.}~\bibnamefont {Fedele}}, \bibinfo {author} {\bibfnamefont {A.}~\bibnamefont {Chatterjee}}, \bibinfo {author} {\bibfnamefont {D.}~\bibnamefont {Reilly}}, \bibinfo {author} {\bibfnamefont {F.}~\bibnamefont {Kuemmeth}}, \bibinfo {author} {\bibfnamefont {M.~F.}\ \bibnamefont {Gonzalez-Zalba}}, \bibinfo {author} {\bibfnamefont {E.}~\bibnamefont {Laird}},\ and\ \bibinfo {author} {\bibfnamefont {N.}~\bibnamefont {Ares}},\ }\bibfield  {title} {\bibinfo {title} {Probing quantum devices with radio-frequency reflectometry},\ }\href@noop {} {\bibfield  {journal} {\bibinfo  {journal} {Applied Physics Reviews}\ }\textbf {\bibinfo {volume} {10}} (\bibinfo {year} {2023})}\BibitemShut {NoStop}%
\bibitem [{\citenamefont {Champain}\ \emph {et~al.}(2023)\citenamefont {Champain}, \citenamefont {Schmitt}, \citenamefont {Bertrand}, \citenamefont {Niebojewski}, \citenamefont {Maurand}, \citenamefont {Jehl}, \citenamefont {Winkelmann}, \citenamefont {Franceschi},\ and\ \citenamefont {Brun}}]{champain2023realtime}%
  \BibitemOpen
  \bibfield  {author} {\bibinfo {author} {\bibfnamefont {V.}~\bibnamefont {Champain}}, \bibinfo {author} {\bibfnamefont {V.}~\bibnamefont {Schmitt}}, \bibinfo {author} {\bibfnamefont {B.}~\bibnamefont {Bertrand}}, \bibinfo {author} {\bibfnamefont {H.}~\bibnamefont {Niebojewski}}, \bibinfo {author} {\bibfnamefont {R.}~\bibnamefont {Maurand}}, \bibinfo {author} {\bibfnamefont {X.}~\bibnamefont {Jehl}}, \bibinfo {author} {\bibfnamefont {C.}~\bibnamefont {Winkelmann}}, \bibinfo {author} {\bibfnamefont {S.~D.}\ \bibnamefont {Franceschi}},\ and\ \bibinfo {author} {\bibfnamefont {B.}~\bibnamefont {Brun}},\ }\href@noop {} {\bibinfo {title} {Real-time milli-kelvin thermometry in a semiconductor qubit architecture}} (\bibinfo {year} {2023}),\ \Eprint {https://arxiv.org/abs/2308.12778} {arXiv:2308.12778 [cond-mat.mes-hall]} \BibitemShut {NoStop}%
\bibitem [{\citenamefont {Chawner}\ \emph {et~al.}(2021)\citenamefont {Chawner}, \citenamefont {Barraud}, \citenamefont {Gonzalez-Zalba}, \citenamefont {Holt}, \citenamefont {Laird}, \citenamefont {Pashkin},\ and\ \citenamefont {Prance}}]{Chawner2021}%
  \BibitemOpen
  \bibfield  {author} {\bibinfo {author} {\bibfnamefont {J.}~\bibnamefont {Chawner}}, \bibinfo {author} {\bibfnamefont {S.}~\bibnamefont {Barraud}}, \bibinfo {author} {\bibfnamefont {M.}~\bibnamefont {Gonzalez-Zalba}}, \bibinfo {author} {\bibfnamefont {S.}~\bibnamefont {Holt}}, \bibinfo {author} {\bibfnamefont {E.}~\bibnamefont {Laird}}, \bibinfo {author} {\bibfnamefont {Y.~A.}\ \bibnamefont {Pashkin}},\ and\ \bibinfo {author} {\bibfnamefont {J.}~\bibnamefont {Prance}},\ }\bibfield  {title} {\bibinfo {title} {Nongalvanic calibration and operation of a quantum dot thermometer},\ }\href {https://doi.org/10.1103/PhysRevApplied.15.034044} {\bibfield  {journal} {\bibinfo  {journal} {Phys. Rev. Appl.}\ }\textbf {\bibinfo {volume} {15}},\ \bibinfo {pages} {034044} (\bibinfo {year} {2021})}\BibitemShut {NoStop}%
\bibitem [{\citenamefont {Peri}\ \emph {et~al.}(2023)\citenamefont {Peri}, \citenamefont {Oakes}, \citenamefont {Cochrane}, \citenamefont {Ford},\ and\ \citenamefont {Gonzalez-Zalba}}]{peri2023beyondadiabatic}%
  \BibitemOpen
  \bibfield  {author} {\bibinfo {author} {\bibfnamefont {L.}~\bibnamefont {Peri}}, \bibinfo {author} {\bibfnamefont {G.~A.}\ \bibnamefont {Oakes}}, \bibinfo {author} {\bibfnamefont {L.}~\bibnamefont {Cochrane}}, \bibinfo {author} {\bibfnamefont {C.~J.~B.}\ \bibnamefont {Ford}},\ and\ \bibinfo {author} {\bibfnamefont {M.~F.}\ \bibnamefont {Gonzalez-Zalba}},\ }\href@noop {} {\bibinfo {title} {Beyond-adiabatic quantum admittance of a semiconductor quantum dot at high frequencies: Rethinking reflectometry as polaron dynamics}} (\bibinfo {year} {2023}),\ \Eprint {https://arxiv.org/abs/2307.16725} {arXiv:2307.16725} \BibitemShut {NoStop}%
\end{thebibliography}%

\end{document}